\documentclass[aps,twocolumn,superscriptaddress,showpacs,preprintnumbers,amsmath,amssymb,amsfonts,floatfix]{revtex4-1}

\usepackage{graphicx,tabularx}
\usepackage{dcolumn}
\usepackage{bm}
\usepackage{color}

\usepackage{upgreek}

\begin{document}


\title{Influence of surface band bending on a narrow  band gap semiconductor: Tunneling atomic force studies of graphite  with Bernal and 
rhombohedral stacking orders}

\author{Regina Ariskina}
\affiliation{Division of Superconductivity and Magnetism, Felix-Bloch Institute for Solid State Physics, University of Leipzig, D-04103 Leipzig, Germany}
\author{Michael Schnedler}\email{m.schnedler@fz-juelich.de}
\affiliation{Peter Gruenberg Institut, Forschungszentrum Juelich,
D-52425 Juelich, Germany}
\author{Pablo D. Esquinazi} \email{esquin@physik.uni-leipzig.de}
\affiliation{Division of Superconductivity and Magnetism, Felix-Bloch Institute for Solid State Physics, University of Leipzig, D-04103 Leipzig, Germany}
\author{Ana Champi}
\affiliation{Centro de Ciencias Naturais e Humanas, Universidade Federal do ABS, Santo Andre, SP-Brasil}
\author{Markus Stiller}
\affiliation{Division of Superconductivity and Magnetism, Felix-Bloch Institute for Solid State Physics, University of Leipzig, D-04103 Leipzig, Germany}
\author{Wolfram Hergert}
\affiliation{Institute of Physics,
Martin Luther University Halle-Wittenberg.
 D-06120 Halle,  Germany}
 \author{R. E. Dunin-Borkowski}
 \affiliation{Peter Gruenberg Institut, Forschungszentrum Juelich,
D-52425 Juelich, Germany}
 \author{Philipp Ebert}
\affiliation{Peter Gruenberg Institut, Forschungszentrum Juelich,
D-52425 Juelich, Germany}
\author{Tom Venus}
\author{Irina Estrela-Lopis}
\affiliation{Institute of Medical Physics and Biophysics, University of Leipzig, D-04107 Leipzig, Germany}

\date{\today}

\begin{abstract}
  Tunneling atomic force microscopy (TUNA) was used at ambient conditions to measure 
the current-voltage ($I$-$V$) characteristics at clean surfaces 
of highly oriented graphite samples with  Bernal and rhombohedral stacking orders. 
The characteristic  curves measured on Bernal-stacked graphite surfaces can be understood with an ordinary self-consistent semiconductor modeling and quantum mechanical tunneling current derivations. We show that the absence of a voltage region without measurable current in the $I$-$V$ spectra is not 
a proof of the lack of an energy band gap. It can  be induced by a surface band bending due to a finite contact potential 
between tip and sample surface. Taking this into account in the model, we succeed to obtain a quantitative agreement between 
simulated and measured tunnel spectra for band gaps  $(12 \ldots 37)$\,meV,  
in agreement to those extracted from the exponential temperature decrease 
of the longitudinal resistance measured in graphite samples with Bernal stacking order. 
In contrast,  the surface of relatively thick graphite samples with rhombohedral stacking reveals the existence of a maximum in the 
first derivative $dI/dV$,  a behavior compatible with the existence of a flat band. The characteristics of this
maximum are comparable to those obtained at low temperatures with similar techniques.  
\end{abstract}

\maketitle


\section{Introduction}
Is ideal graphite, a carbon-based structure composed by weakly  coupled graphene layers, a semimetal or a narrow-gap semiconductor? This fundamental question and the possibility of a  spontaneous symmetry breaking\cite{zha10} that may trigger a narrow energy gap, was not clarified  in earlier experiments. The main reason is that 
 most of the earlier experimental studies were done using samples with a considerable amount of highly conducting stacking 
 faults (SF) parallel to the graphene planes of the
 graphite structure \cite{gar12}. 
 The dominant stacking order of the graphene layers is the  
Bernal (2H) stacking order (ABABA...). There is also a minority phase, called  rhombohedral (3R)  (ABCABCA...), occurring  with a concentration  $\lesssim 25\%$ in bulk samples \cite{kelly,lin12,pre16}. 
Hence, well ordered graphite samples  contain   SF being
  two dimensional (2D) interfaces between twisted 2H, twisted 3R regions or  between the 2H and 
 3R stacking orders. Several reports in the last
  12 years on the 
  internal structure of well ordered, pyrolytic as well as natural graphite samples  characterized by transmission electron microscopy (TEM)
  and X-ray diffraction (XRD),   revealed a significant amount
  of SF separated by  a few tens to several hundreds of nm in the $c-$axis direction \cite{bar08,gar12,bal13,schcar,bal15,pre16,chap7}.  These 2D SF show quite different electronic properties, which dominate the conductance  at  certain temperature and magnetic field ranges of high quality, highly ordered graphite samples\cite{zor18,pre19}.  
  
  Scanning tunneling microscopy (STM) measurements done on well ordered graphite or bilayer graphene samples at room and low temperatures, indicate that those SF can be also found at or very near the sample surface\cite{kuw90,flo13,mil10} altering the  density of  states (DOS) locally.\cite{bri12}  Kelvin force microscopy studies of the surface of well-oriented graphite samples,  done in air as well as in  inert atmosphere, revealed the coexistence of insulating- and conducting-like regions,\cite{lu06} whose origin can  also be related to the presence of ideal graphite and regions with SF at or near the surface, respectively. These highly conducting 2D SF are the origin for the metallic-like behavior in the temperature dependence of the resistance\cite{gar12,zor18},  
 for the low temperature 
  Shubnikov-de Haas and de Haas-van Alphen quantum oscillations  \cite{chap7,zor18}, and  also for the huge diamagnetism of graphite\cite{mcclure56} at fields applied parallel to the $c-$axis\cite{sem18}.  In other words, the proposed Fermi surface \cite{kelly} does not correspond to ideal graphite, in contradiction to the
  semimetal picture proposed more than 60 years ago.\cite{mcc1,sw58,mcc2} Finally, some of the SF are at the origin of  the observed granular superconducting behavior of graphite: 
   whereas twisted graphene bilayers show superconductivity at $T < 10~$K\cite{cao18} (related to the existence of a flat band \cite{mar18}),   higher critical 
  temperatures have been reported earlier due to internal and larger SF in bulk and TEM lamellae\cite{bal13,schcar,bal14I,bal15},  
  partially containing the 3R phase\cite{pre16,esq18}.

The experimental facts that speak for a semiconducting nature of ideal graphite are the following.\\ {\em -Longitudinal electrical resistance}:  Thin graphite samples with low or negligible amount of SF
show an exponential temperature dependence in  the electrical resistance compatible with a semiconducting behavior\cite{gar12} with an energy gap
in the order of $\sim 30 \pm 10~$meV for the 2H phase and $\sim 110 \pm 15~$meV for the 3R phase\cite{zor17}. 
 The small band gaps of the 2H or 3R phases combined with the huge electrical anisotropy  of  graphite,
 as well as the contribution of the 2D SF to the total conductance of a sample \cite{gar12} can be anticipated to be at  
 the origin of the
complex, even contradictory temperature and magnetic field  dependences of the transport properties 
found in literature.\cite{zor18,chap7} An exponential decay with temperature  of the longitudinal resistance can be taken as a  semiconducting 
fingerprint that  challenges the usual semimetal description of the band structure of ideal graphite,
assumed in the past on the basis of the McClure, Slonczewski and Weiss calculations\cite{mcc1,sw58,mcc2}. 

{\em -Magnetoresistance}: Indirect support for the semiconducting behavior of graphite is given by the magnetic 
field dependence of the magnetoresistance 
 at  temperatures  $T$ above 50~K and fields to 65~T. In these ranges, the contribution of the SF  to the magnetoresistance 
 turns out to be negligible in comparison to that of 
 the graphene matrix\cite{bar19,pre19}. The field dependence of the magnetoresistance  can be well understood  within a two-band model indicating the
existence of an energy gap between a valence and a conduction band\cite{bar19}.  
 
{\em -Hall effect}: A further  example of the influence of defects in graphite samples is the sign of the Hall coefficient.
Out of thirteen published studies on the Hall coefficient  of different graphite samples (not few-layers graphene) \cite{kin53,mro56,sou58,coo70,bra74,osh82,bun05,van11,yakovprl03,kempa06,kopepl06,sch09,esq14}, nine  reported
positive Hall coefficient at a certain field and temperature range\cite{kin53,mro56,sou58,coo70,bra74,osh82,bun05,van11,esq14}. The differences in the Hall coefficient 
have been partially explained  by  taking into account  SF within the graphite matrix\cite{esq14}. 
Graphite appears to be another  example of solids for which both, intrinsic and extrinsic effects 
 contribute to the Hall coefficient, making a direct estimate of the intrinsic carrier 
densities difficult\cite{esq14}. 

 {\em -Optical spectroscopy}: Optical pump-probe spectroscopy with 7-fs pump pulses  indicates 
that at ultrafast time scale graphite does not behave as a semimetal but as a semiconductor\cite{bre09}.

\begin{figure}
 \includegraphics[width=1.05\columnwidth]{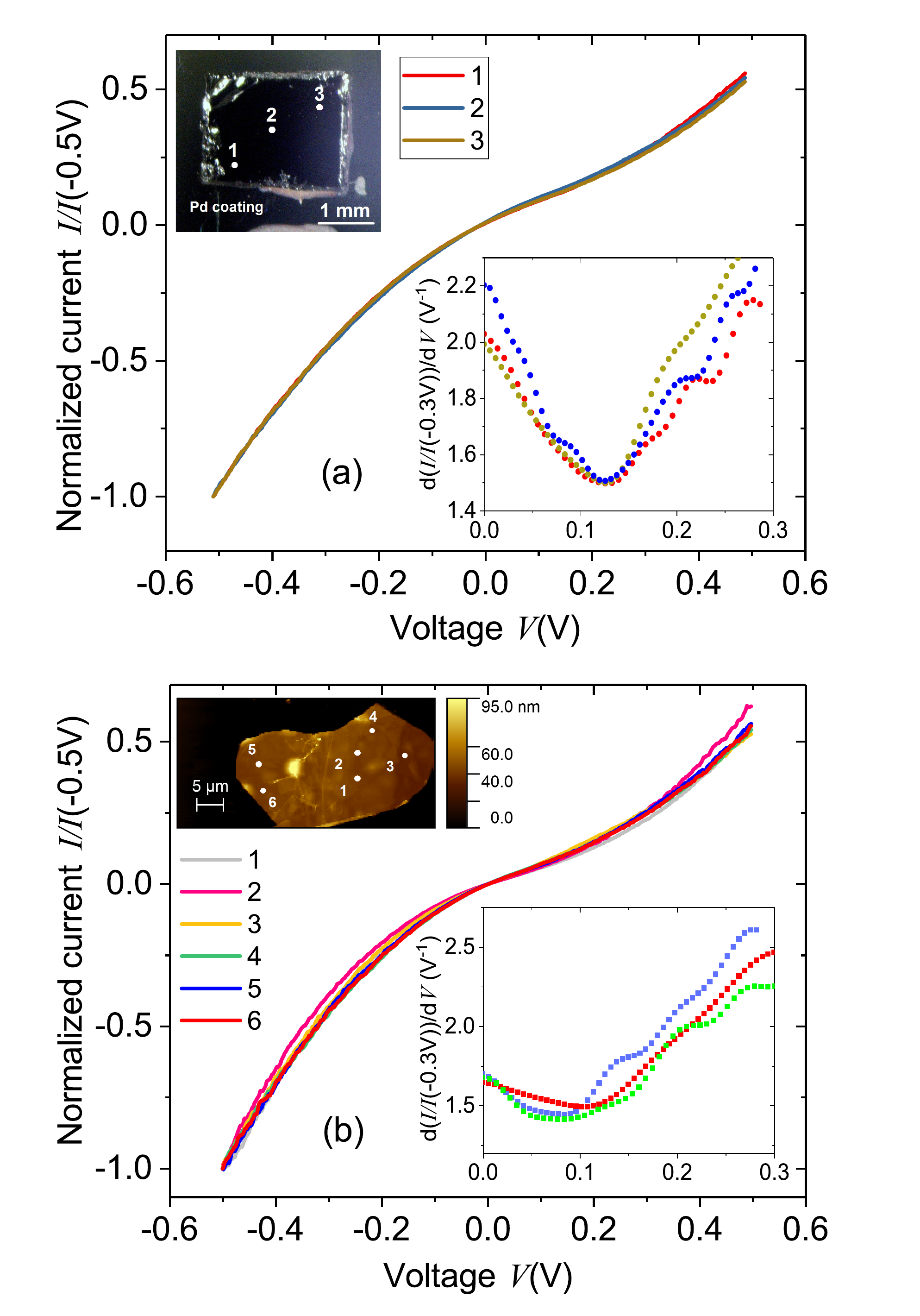}  
  \caption{(a) Normalized TUNA current-voltage characteristics measured in the tunneling regime at different positions 
  of a HOPG sample (HOPGA-Pd-1, see Table~\ref{samples}), see optical image in the upper left inset.  
  The maximum current values at -0.5~V~used for the normalization were (1$^{\text st}$ to 3$^{\text rd}$ locations): 217~pA, 302~pA, 240~pA.
  The lower right inset shows the first derivative of the $I$-$V$ curves at the three locations. The normalization of the
  current values  with the first derivative was done at -0.3~V~(instead of -0.5~V~as in the main panel) to enhance the details
  in the region of interest. Figure~\ref{dlc} shows the first derivative in the whole measured voltage range for the same sample. 
  (b) Similar measurements for a graphite mesoscopic 
  flake (sample MG-Pd-11-2) (see SEM image at the upper left inset) at different positions. The maximum current values at -0.5~V~used for normalization
  were: 1174~pA, 912~pA, 565~pA, 597~pA, 671~pA, 447~pA.   
  The inset below right shows the first derivative of the $I$-$V$ curves at three locations (current normalized
  at -0.3~V). All results were obtained at ambient conditions.}
  \label{gra}
\end{figure}

Regarding studies on graphite with the two different stacking orders 
using angle-resolved photoemission spectroscopy (ARPES)  we
would like to mention the following recent publications.   A gaplike feature of 25~meV between the $\pi$ and $\pi^\star$  
bands at the K(H) point was reported \cite{sug07}, whereas in \cite{gru08}  an energy gap of 37~meV 
was inferred from the band fits near the H point. A gaplike feature at an energy of $\simeq 67~$meV was
reported in \cite{liu10} and interpreted as  a phonon-induced gap. Recently, similar results were obtained by
ARPES in combination with scanning tunneling microscopy/spectroscopy (STM/STS) studies in highly oriented pyrolytic 
graphite (HOPG) of, however, unreported quality \cite{yin20}.   Very similar 
$dI/dV$ characteristics curves with clear gaplike features were obtained by STS, 
independently of the used tip material (Ag, Pt/Ir or W), prior to
exposing the sample to hydrogen molecules \cite{yin20}.  
This similarity is actually not expected because of the large differences
between the contact potentials of the used tip materials and graphite, a fact that affects the $I-V$ characteristics, as we will see
below. Improvements in the ARPES technique allow measurements 
with a resolution of $\lesssim 100~$nm in the sample surface plane. 
This is a necessary resolution because of the lack of large single crystalline regions in usual graphite samples \cite{gon07}. 
We note, however, that the usual energy resolution of $\gtrsim 5~$meV appears still not enough  to clearly resolve a 
semiconducting energy gap of the order of 30~meV. Nano-ARPES measurements on a long sequence of 3R stacking order
showed the existence of a  flat band, which 25~meV dispersion appears to be 
compatible with a magnetic ground state characterized by an energy band gap close to 40 meV \cite{hen18}.

 The aim of this work is to unravel whether tunneling atomic force spectroscopy
  can provide further evidence for  the semiconducting nature of graphite and whether
  differences can be measured at room temperature between the two stacking order phases. 
  We used a 
  relatively new tunneling spectroscopy technique, called  tunneling atomic force microscopy or TUNA, 
   and  in particular the PeakForce operation mode \cite{PF}. 
   The overall results are compatible with the interpretation that ideal graphite with 2H phase is
a narrow band-gap semiconductor and that the 3R phase, found in thick flakes, exhibits a  maximum  in the DOS at its surface,
compatible with the existence of a flat band in the electronic spectrum.\cite{pie15,kop13,vol18,hen18} 
  
\section{Results and discussion}

All TUNA results presented in this work were obtained in air at normal conditions on natural and HOPG samples of grade A and micrometer size flakes obtained from the corresponding bulk pieces. The quality and
stacking order characterization of the samples were done using Raman spectroscopy, as done in previous reports\cite{con11,hen16,tor17,ram20}. 
The $I$-$V$ curves  were obtained in the
tunneling regime  with currents between a few tens of pA to  $\lesssim 3~$nA .  Further details about samples 
and characterization methods are provided in the `Samples and Methods" Section~\ref{sm} below.

\subsection{Bernal stacking order}
\subsubsection{Experimental tunneling spectra}

 Figure~\ref{gra} shows the normalized $I$-$V$ characteristics and their first derivatives (see insets) for the bulk HOPG sample (a)
and a graphite flake (b). Raman spectroscopy indicates that the main phase in those samples is the 2H phase. 
Figure~\ref{gra}(a) illustrates normalized $I$-$V$ curves of the bulk samples, measured at three different spatial 
positions. All three curves are nearly  identical. Analogous  measurements on the thin graphite flakes 
reveal similar $I-V$ curves as for the bulk sample, although with somewhat larger  local variations, as depicted in Fig.~\ref{gra}(b). 
These small spatial  variations of the current can be observed in general for thin graphite flakes 
and are not necessarily related to the presence of SF at the surface but due to bending or defective regions.

Our normalized $I$-$V$ curves are asymmetric and  
the first derivative of the normalized current exhibits a minimum 
at $V \simeq$ 0.12~V, see the inset in Fig.~\ref{gra}(a).  
An early STM work on graphite at ambient conditions reported 
a relatively wide minimum at low voltages in the differential conductance.\cite{agr92}  
Similar curves were obtained  at 4.2~K on very thin graphene 
multilayer samples with the Bernal phase\cite{pie15}.

At first view, the normalized $I$-$V$ curves in Fig.~\ref{gra} exhibit  apparently no band gap, i.e. a voltage region 
without detectable tunneling current.  On the other hand, the  first derivative $dI/dV$ 
is a quantity generally assumed to be proportional to the LDOS.
Although in a classical semiconductor model 
a minimum in the LDOS can be a trace of a band gap, it is not a sufficient 
criteria to infer its existence. The simulation of the $I-V$ spectra proposed below can help
to discern to which extent the existence of a (small) energy gap is compatible with the measured
spectra.

\subsubsection{Simulation of the spectra}
In view of the aforementioned difficulties that prevent a direct observation  
of a band gap using the $I$-$V$ curves, we turn to a different approach: First, 
we assume that graphite behaves like an ordinary semiconductor. Under this assumption,
 a software package, specifically designed for simulating tunneling currents at 
semiconductor surfaces,\cite{schnedler:2015b,schnedler:2016a} is employed to investigate the influence of the size of 
the band gap on the $I$-$V$ curves. By comparing simulated $I$-$V$ curves with those 
experimentally obtained on graphite  with Bernal stacking, we conclude that 
the measured spectra can be understood using an ordinary semiconductor model 
by assuming small band gaps. Finally we try to estimate the size of the band
 gap from the TUNA measurements.

In order to derive the tunneling current, we follow the 
two-step method as described in Refs.\,\cite{schnedler:2015b,schnedler:2016a}~: 
First, we performed self-consistent electrostatic simulations of the 
tip-vacuum-semiconductor system to unravel the potential- and carrier 
distributions in three dimensions. A Pt-Ir tip with a radius of 100\,nm 
and opening angle of 45\,$^\circ$ was chosen. The carrier concentrations of 
the sample are derived using the parabolic band approximation (see, e,g, Ref.~\cite{roz07}).  Band gaps 
ranging from 0.1\,meV to 120\,meV are assumed. The graphite's Fermi-level 
position $E_\text{F}$ as well as the concentration of free 
electrons $n\left(E_\text{F},T\right)$ and holes $p\left(E_\text{F},T\right)$ 
at a temperature of $T=300$\,K are defined by solving the charge neutrality condition:
\begin{equation}
 n\left(E_\text{F},T\right)-p\left(E_\text{F},T\right) + N_\text{A}\left(E_\text{F},T\right)-N_\text{D}\left(E_\text{F},T\right)=0\,, 
 \label{neu}
 \end{equation}
 with $N_\text{A}\left(E_\text{F},T\right)$ and $N_\text{D}\left(E_\text{F},T\right)$ 
being the acceptor and donor concentrations. Several Hall effect 
results\cite{kin53,mro56,sou58,coo70,bra74,osh82,bun05,van11,esq14} 
suggest that graphite should be degenerated with a Fermi-level 
located inside the valence band with 
a free carrier concentration of the order of $10^{17}$cm$^{-3}$ at 300~K.\cite{esq14,dus11} 
We assume density of states effective masses of the order of one hundredth of the electron rest mass 
for a parabolic band approximation. 
Using these constraints, we can solve the charge neutrality condition only, 
if a certain concentration of shallow acceptor states are incorporated. 
Due to the lack of the precise knowledge about the graphite density of states 
effective masses and acceptor ionization energies, it remains to be clarified in a future work, whether the assumed 
concentration of shallow acceptors states depends  on the  concentration of atomic lattice defects in  our HOPG samples.
Note that  those defects can play a role in the effective carrier density, as irradiation studies indicate.\cite{arn09} 

Without bias ($V=0$), the contact potential is defined as the work function difference between the metallic probe tip (Pt$_{80}$-Ir$_{20}$) and the graphite sample. It affects the tip-induced band bending in the same manner like the application of a bias voltage.  For Pt and Ir work function values between 5.7 and 5.8\,eV were reported\cite{kaa95}.  On the other hand for graphite the work function varies from 4.6\cite{krishnan:1952} to 4.7\,eV\cite{rutkov:2020}. Taking into account these values, the contact potential between the probe tip and graphite surface is expected to be $\simeq 1$\,eV.

In a second step, we use the one dimensional electrostatic potential 
along the central axis through the tip apex to derive the tunneling 
currents through the vacuum barrier using the WKB approximation based
 model described in Refs.\cite{bono:1986,feenstra:1987b}\,.
In order to fit the measured $I$-$V$ curves to the  calculated ones, 
we took the band gap, the effective masses, the concentration of acceptors and  
the tip-sample separation as fit parameters.
\begin{figure*}
		\centering
		\includegraphics[width=0.85\textwidth]{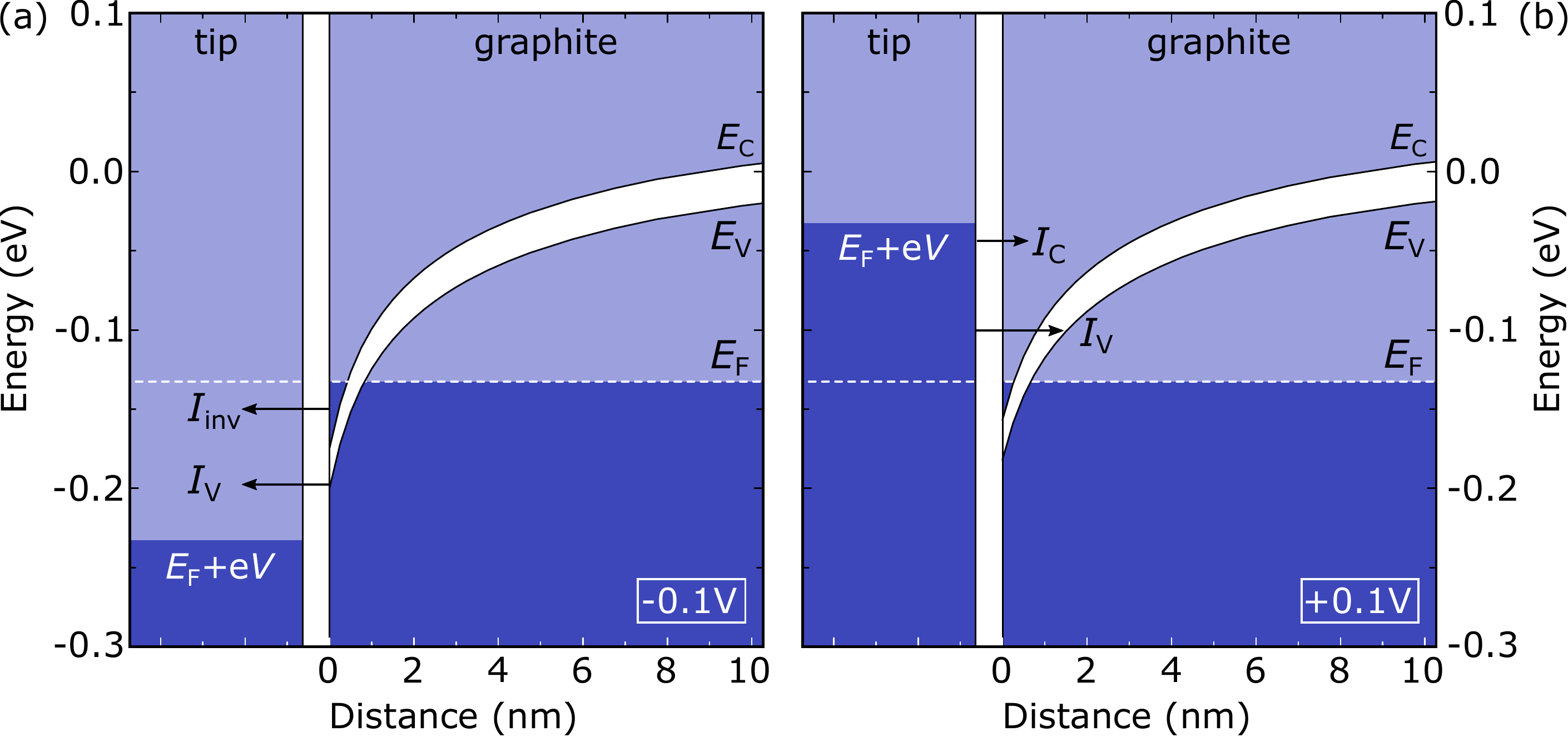}
				\caption{Tip-induced band bending as a function of the 
distance from the graphite surface for a biased tip-vacuum-sample system. 
(a) At negative sample voltages, tunneling out of filled valence band states ($I_\text{V}$) 
is dominating the total tunneling current. In addition, due to the large downward 
band bending and the small band gap, the conduction band minimum is dragged below the 
Fermi level at the surface, filling up its states with electrons. These electrons can 
tunnel out of the conduction band into empty tip states, leading to an additional, but 
small, tunneling current component ($I_\text{inv}$). (b) The large downward band bending 
is also present at positive sample voltages, due to a large contact potential of $\simeq 1$\,eV. 
The total tunneling current is again composed of two components: The first component 
is driven by tunneling of electrons from the tip into empty valence band states ($I_\text{V}$). 
This is only possible because the Fermi-level of the sample is situated below the valence 
band maximum, leading to unoccupied states within the valence band. The second component is 
driven by electrons tunneling from filled tip states into empty conduction band states ($I_\text{C}$). 
At small positive sample voltages, $I_\text{V}$ is the dominant tunneling current component 
while for large positive voltages, $I_\text{C}$ becomes the largest contribution to the total 
tunneling current.}
		\label{bandbending}
	\end{figure*}

\subsubsection{Modeling of the tunneling current based on the tip-induced band bending} 
\begin{figure*}
		\centering
		\includegraphics[width=0.9\textwidth]{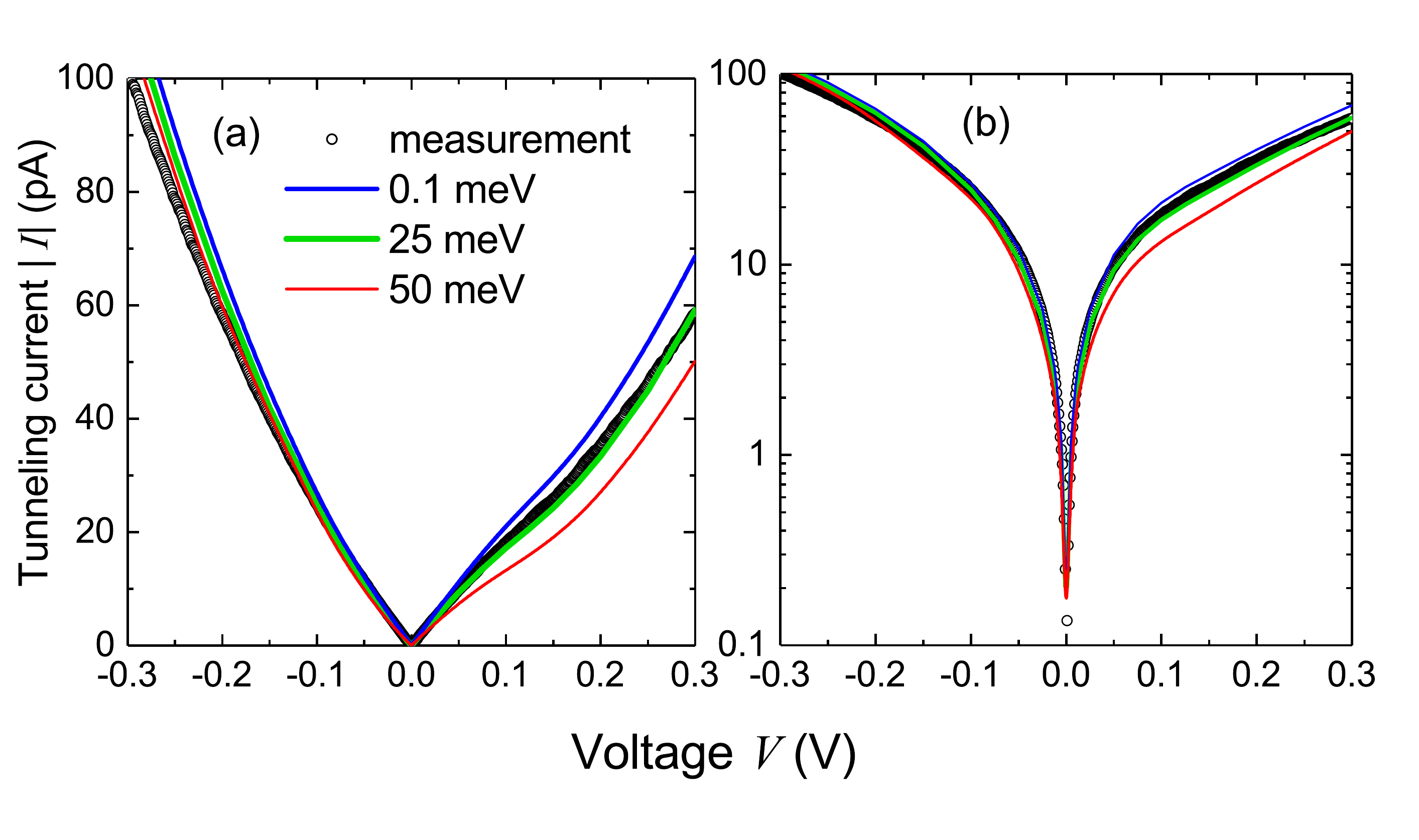}
		\caption{Comparison of TUNA spectra measured at clean and atomically 
flat Bernal-stacked graphite surfaces at $T=300$\,K. The symbols correspond to the
		curve obtained at position 1 for the HOPG sample (HOPGA-Pd-1) 
shown in Fig.~\ref{gra}(a) with the best fit tunneling current simulation (green solid line). The blue solid line 
was obtained  choosing a  band gap of 0.1~meV with the same set of parameters, see Fig.~\ref{bandgapseries}.
The results are plotted in a (a) linear and (b) logarithmic scale. In order to better recognize the asymmetry of the
$I$-$V$ curves between the positive and negative voltage ranges, we plotted the absolute values of the tunneling current values of the experimental 
as well as the computed currents. Note the change in the curvature of the $I$-$V$ curve at $V \sim 0.12~$V and
the difference in the absolute values of the current at $\pm 0.3$~V.}
		\label{simresults}
\end{figure*}
As discussed above, from experiments we expect a  free carrier concentration of the order 
of  $\sim 10^{17}$\,cm$^{-3}$ (i.e. $\sim 3 \times 10^9$~cm$^{-2}$ per graphene layer) at 300~K for  Bernal graphite. 
For such a carrier concentration the electrostatic potential, present between the metallic probe 
tip and graphite during STM experiments, cannot be screened completely at the graphite 
surface and  can reach  the subsurface region of the material 
(commonly referred to as tip-induced band bending). 
We take the tip-induced band banding into account in our self-consistent simulations.

The above discussed contact potential of $\simeq 1\,$eV in conjunction with small band gaps lead to empty and filled states, which are simultaneously present in \textit{both}, valence- and conduction-band. Hence, at negative bias (sample) 
voltages the tunneling out of filled valence- and conduction-band states into empty tip states occurs, while at positive voltages, tunneling from filled tip states into empty valence- and conduction-band states takes place. This situation is illustrated in 
Fig.~\ref{bandbending}.  Even at positive sample voltages, the downward band bending remains present. In contrast to other $p$-type semiconductors with larger band gaps, tunneling into empty conduction band states ($I_\text{C}$) occurs even for very small positive sample voltages already (in addition to the tunneling current $I_\text{V}$ into empty valence band states).  At a sample voltage of approximately $\sim +0.12$~V the increasing $I_\text{C}$ becomes larger than the decreasing  $I_\text{V}$,  leading to the above discussed minimum in the first derivative $dI/dV$.

A good agreement between the measured and simulated $I$-$V$ curves at Bernal stacked graphite surfaces was achieved as indicated by the green solid line in Fig.~\ref{simresults}. This simulation was obtained for a free carrier concentration of $2.3 \times 10^{17}$~cm$^{-3}$, similar to the expected one\cite{esq14,dus11}, a band gap of 25\,meV, similar to that obtained from transport results\cite{gar12,zor17}, as well as effective masses of 0.01 and 0.0075 for the valence- and conduction-band, respectively. We restricted the calculations within the voltage range of interest  $\pm 0.3~$V. In the same figure we show the simulations obtained with the same parameters but for band gaps 0.1~meV and 50~meV. 
\begin{figure}[]
		\centering
		\includegraphics[width=0.5\textwidth]{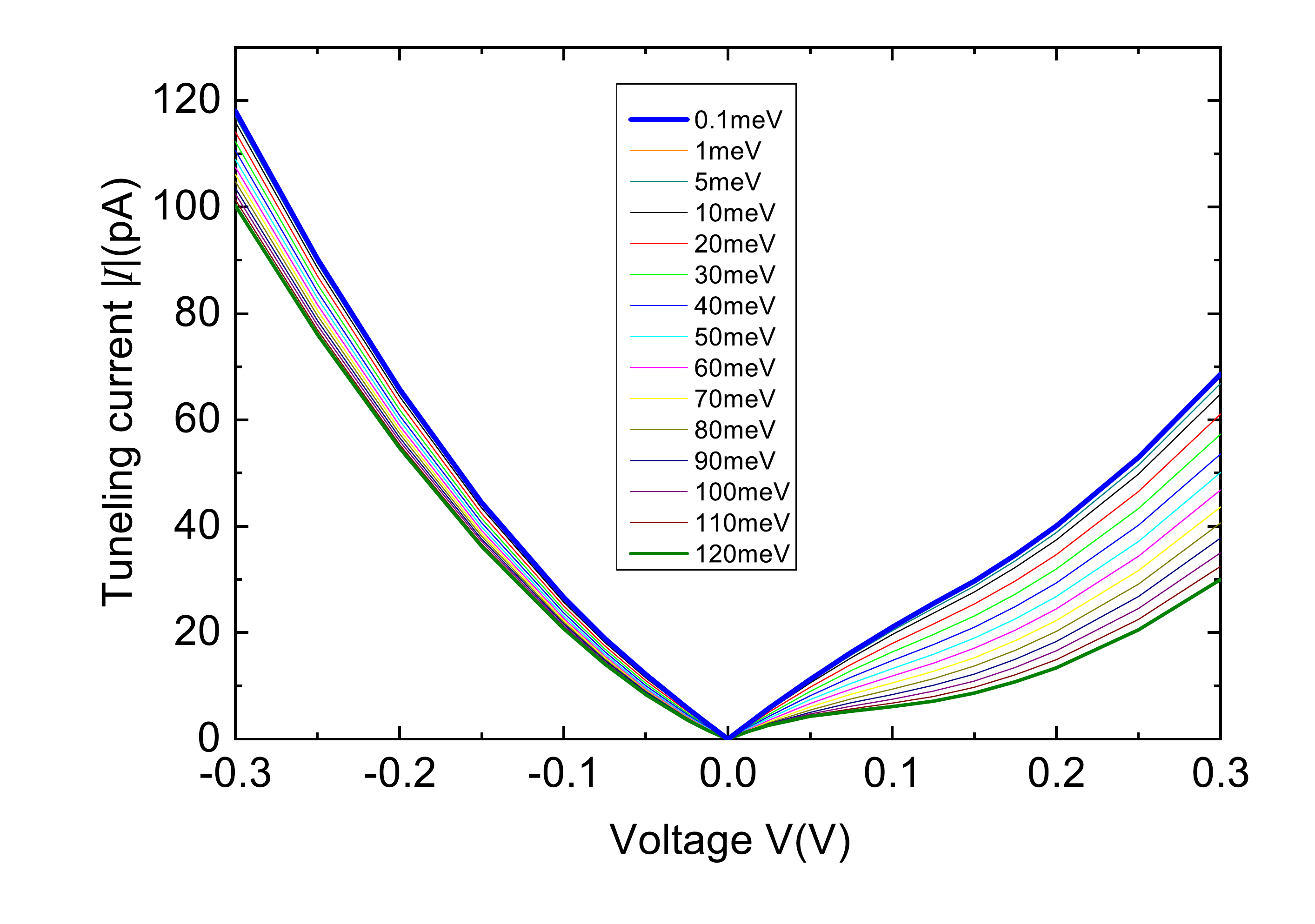}
		\caption{Simulated tunneling current vs. bias voltage at different band gaps between 0.1\,meV and 120\,meV using the parameters explained in the main  text. Note
		that the absolute values of the current are plotted.}
		\label{bandgapseries}
\end{figure}

In order to investigate the influence of a band gap  on the tunneling spectra of graphite, we calculated $I$-$V$ curves for band gaps between 0.1\,meV and 120\,meV. 
Figure \ref{bandgapseries} shows that 
in particular the tunneling current at the positive voltage branch decreases with increasing band gap. This behavior 
is caused by an interplay of two effects: First, the shift of the graphite's conduction band states towards larger 
energies with increasing band gap leads to less conduction band states that are available for tunneling. Second, 
a larger band gap leads to an energetically increased tunneling barrier between tip and sample, which in return 
decreases the tunneling probability. Hence, also the tunneling current  related to valence band states decreases slightly 
with increasing band gap. This delicate interplay alters the slope of the combined tunneling current (i.e. the sum of conduction- and 
valence-band related tunneling currents) at positive sample voltages and defines the position of the minimum 
of the $dI/dV$ curves. Larger band gaps result in stronger curvatures at the positive voltage branch, while for 
small band gaps the positive voltage branch becomes more linear. In the limit of a vanishing band gap, the 
minimum in the $dI/dV$ curve is still present. Note that  thermal broadening of the Fermi-Dirac distribution of the Pt$_{80}$Ir$_{20}$-tip states is not taken into account in the tunneling current computation, which can however be anticipated to not change the position
of the minimum in the $dI/dV$ curve but only its intensity. 

\begin{figure}
		\centering
		\includegraphics[width=0.5\textwidth]{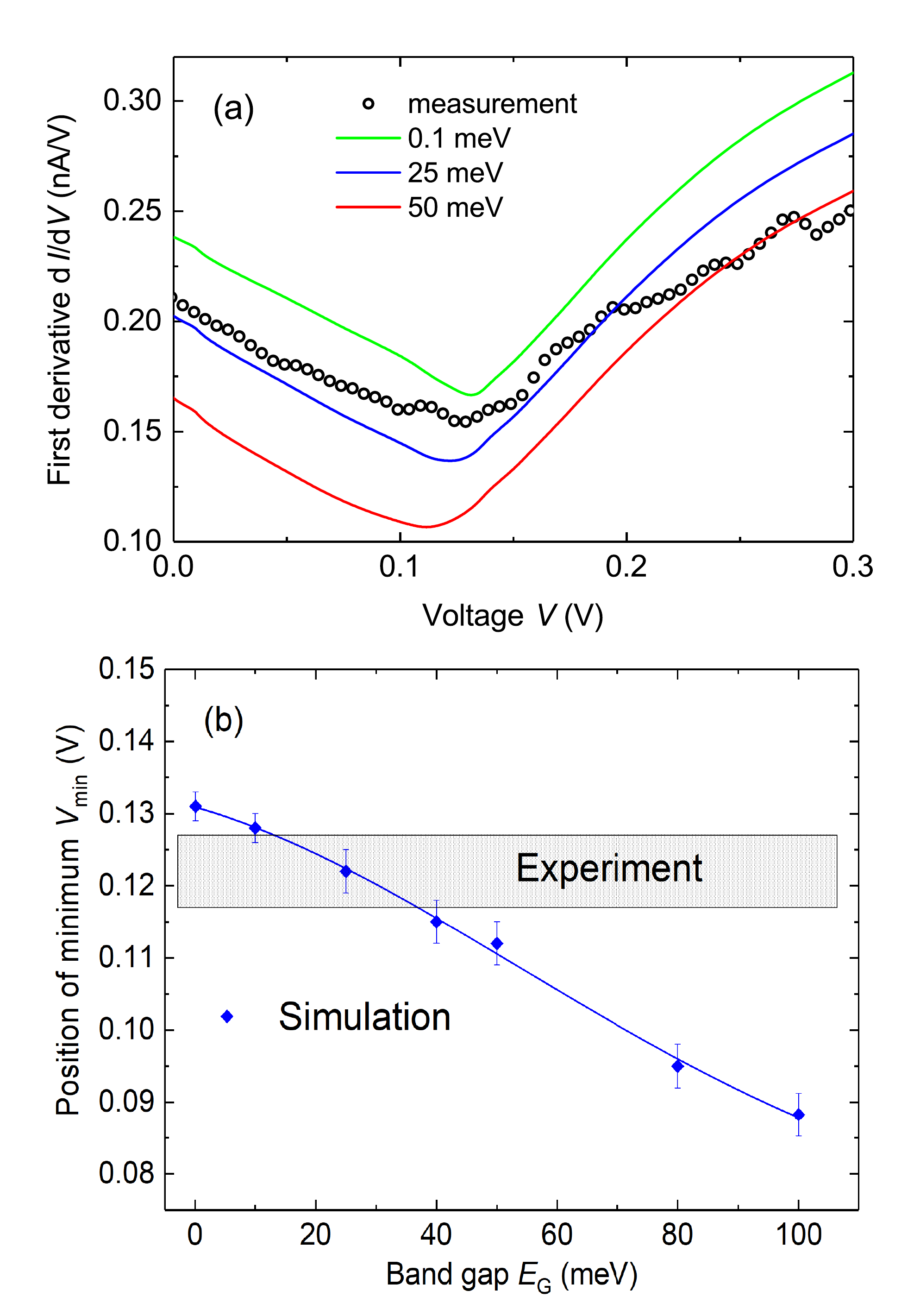}
		\caption{(a) $(\circ)$: First derivative of the tunnel current  vs. applied voltage measured on the HOPG-Pd-1 sample shown in Fig.~\ref{gra}(a)   and 
		(lines) of the simulated tunnel currents at three different band gaps (0.1, 25, 50)~meV with otherwise unchanged parameters.
		Note that the absolute values of the derivative are plotted. 
		(b) Voltage position of the minimum of the first derivative vs. the band gap $E_G$. The ($\blacklozenge$)  were obtained from the simulations. The  
		line is only a guide to the eye. The range of the
		minimum in the first derivative of the experimental $I-V$ curves of the HOPG-1 sample is given by the shadowed rectangular region. }
		\label{der1}
\end{figure}

Although the simulations with smaller and larger band gap appear to agree less with the experimental data according to the
depicted graph $|I|$ vs. $V$ shown in Fig.~\ref{simresults}, a clear distinction is difficult on this basis alone. A 
more sensitive way is given by comparing the first derivative and in particular the position of its minimum. 
The reason is that  the minimum in the $dI/dV$ curve 
is found to shift with the band gap as illustrated in Figs.~\ref{der1}(a) and (b). 
As one can recognize in this graph, the  first derivative of one of the $I-V$ curves obtained for sample
HOPG-Pd-1 as example (see Fig.~\ref{gra}(a)),  roughly agrees with the simulation with energy gap of 25~meV with
a minimum  at $V_{\text {min}} = 0.127 \pm 0.007$~mV. 
We further quantified the voltage position of the experimentally observed minima   by fitting a polynomial  of 9th grade to the numerical differentiation 
of all $I-V$ curves, yielding an average value of $V_{\text {min}} = 0.122 \pm 0.005$~mV. The experimental region of the
minimum is shown as shadowed area in Fig.~\ref{der1}(b). The comparison of the measured $V_{\text {min}}$ values with the simulated 
$V_{\text {min}}$ vs. the band gap $E_G$  suggests that the experimental data can be described best with a
 semiconductor model with a band gap between 12~meV and  37~meV. 
 From the simulated  $V_{\text {min}}(E_G)$, see Fig.~\ref{der1}(b), we note that the data obtained for the
 mesoscopic graphite sample with Bernal stacking order, see Fig.\ref{gra}(b), suggest that 
 larger and a broader distribution of band gaps 
 could be localized at certain regions of the inhomogeneous or bended graphite surface. We may speculate that atomic lattice 
 defects, other than  simple  SF, but twisted or turbostratic stacking  with more complexes sequences 
 may have different band  gaps.

We varied the parameters of the simulations to  estimate the error range. A decrease of 
the contact potential leads to a slight increase of the band gap ($\sim 4~$meV per 0.1~eV decrease) 
while decreasing the  free carrier concentration ($\sim 4 \times 10^{16}$~cm$^{-3}$ per 0.1~eV decrease) 
and  the effective band masses.  Taking into account transport data,\cite{esq14,dus11} we expect that 
the carrier concentration  is at least $\gtrsim 10^{17}~$cm$^{-3}$ at 300~K. Hence, the contact potential should be $\gtrsim 0.7~$eV, which restricts the 
band gap to $\lesssim 40~$meV.  Taking into account correlation effects between the parameters, we
estimate  an error in the carrier density parameter of $\pm 1.5 \times 10^{-17} $cm$^{-3}$. 
We note that if we assume a negligible band gap at contact potentials smaller than 1~eV, the deviation between the measured 
data and the theoretical curve increases, pointing further  to the existence of a finite energy gap.

  We conclude this section by pointing out that
the measured spectra for the Bernal stacking order are 
best described by a semiconductor model with a narrow band gap. In first approximation the first derivative
spectra is proportional to the LDOS, although the voltage scale is shifted by the existence of a tip-induced
band bending. Taking advantage of these insights in the interpretation of the tunneling spectra on Bernal stacking
order graphite, we now turn to measurements of the  rhombohedral stacking order of graphite.

\subsection{Rhombohedral stacking order}

\begin{figure}[h]
 \includegraphics[width=1.15\columnwidth]{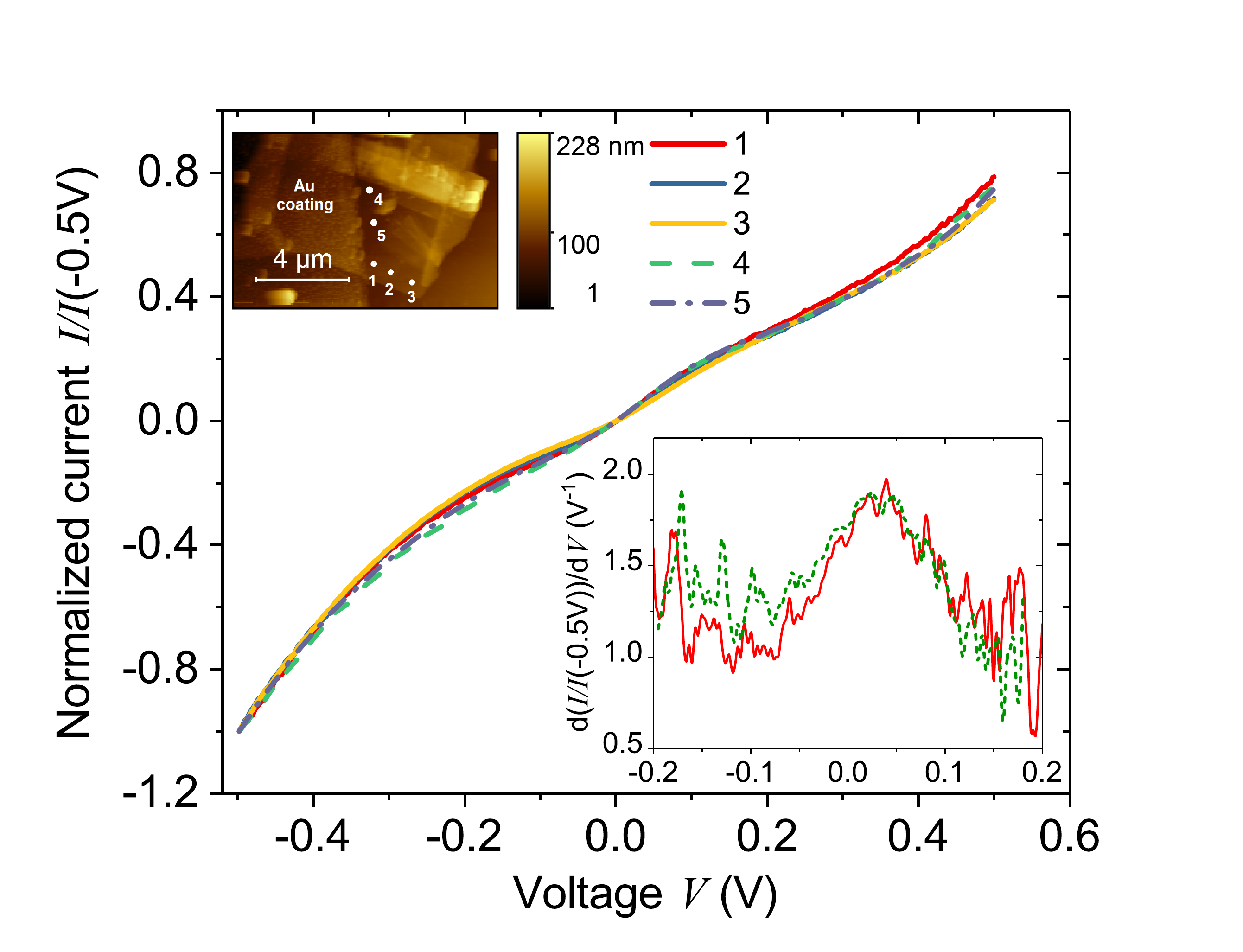}  
  \caption{TUNA current-voltage characteristics obtained at different positions of the 3R phase surface of a 
   22~nm thick ($\simeq 22$ unit cells) natural graphite flake (sample NBF5-01-03, see Table I), see upper left inset with the AFM image.   
  The   current values used for normalization were (from 1$^{\text {st}}$ to 5$^{\text {th}}$ positions): 276~pA, 293~pA, 386~pA, 374~pA, 366~pA. The bottom
  right inset shows the first derivative of the $I$-$V$ curves at the 1$^{\text {st}}$ and 4$^{\text {th}}$ positions in the voltage region of interest. }
  \label{3R}
\end{figure}

\begin{figure}[h]
 \includegraphics[width=1.1\columnwidth]{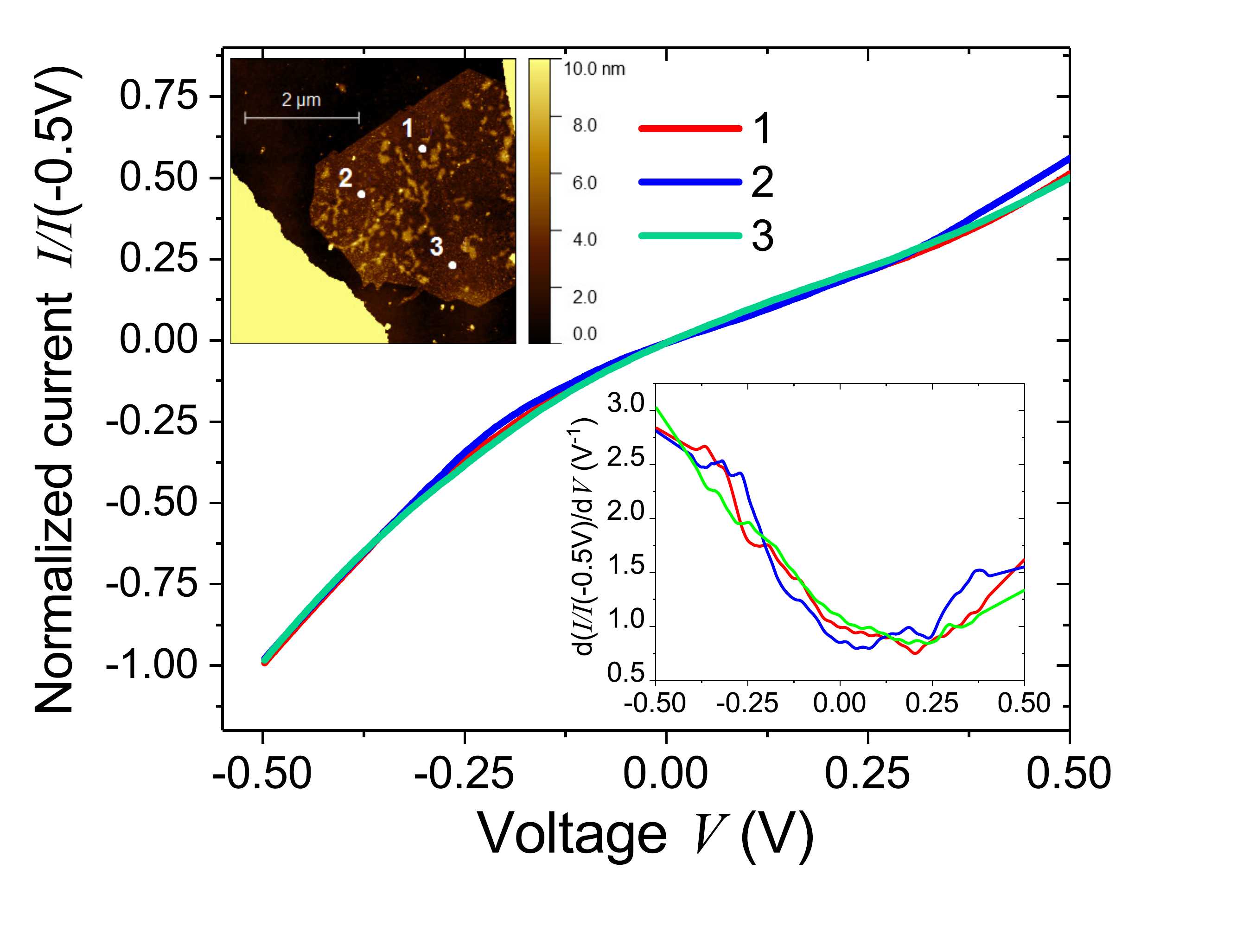}  
  \caption{TUNA current-voltage characteristics  obtained at three different positions of the 3R phase surface of a 3~nm thick  
   ($\simeq 3$ unit cells) natural graphite flake (sample NBF5-SNG-3, see Table $I$), see upper left inset with the AFM image.   
  The   current values at -0.5~V used for the normalization were (from 1$^{\text st}$ to 3$^{\text rd}$ positions): 241~pA, 222~pA, 213~pA. 
  The bottom right inset shows the first derivative of the normalized current $I(V)$ curves. }
  \label{3R-thin}
\end{figure}

The 3R phase is of
  especial interest nowadays due to the expected flat band at its surface or at its interfaces with the 2H stacking order\cite{kop13,mun13}. 
  A flat band in the 3R phase has been found experimentally\cite{xu15,pie15,hen18,wan18}, which correlates with a maximum 
  in the DOS at the Fermi level, enhancing 
  the probability to trigger superconductivity \cite{kop13,hei16,vol18} 
and/or magnetism\cite{pam17,hen18,oja18} at high temperatures. 
 In particular, STS obtained on a sequence of five layers of 3R phase showed
a peak in the DOS around the Fermi level with a width of $\sim 50~$meV at half maximum\cite{pie15}. We have performed TUNA measurements in several natural graphite samples with the 3R phase. For a relatively thick (22~nm) sample with the 3R phase, we measured a clear maximum in the differential conductance reproducible at
different positions spread by several $\mu$m$^2$ in the sample area, see Fig.~\ref{3R}. This maximum is not observed, however, in a
3R phase sample of much smaller thickness (3~nm), see Fig.~\ref{3R-thin}. This difference may indicate that either the 
roughness of the samples plays a detrimental role  or the number of 3R unit cells in the thin sample is not enough
to clearly develop this feature at room temperature, as numerical simulations suggest.\cite{kop13}

We note that the maximum in the differential conductance observed in the thicker 3R sample is shifted by $\sim 40~$meV above the zero level and  with a width at half maximum of $\sim 100$~meV. Those values are of the same order
as reported at 4.2~K, in spite of  $\simeq  70$ times higher temperatures, stressing the robustness of the high DOS feature
around the Fermi level at the surface of this graphite phase. The shift of the maximum in the differential conductance may come from
the influence of a finite contact potential and band bending.

We note that the temperature dependence of the resistance measured for  a large number of bulk samples from different origins follows over a very broad range (2~K~$\lesssim T \lesssim 1100$~K) the temperature dependent resistance known for semiconductors, suggesting, that this minority 3R phase should also be a narrow gap semiconductor with an energy gap of the order of 100~meV\cite{zor17}. Under this assumption we would have the situation of a different band structure at the surface from that in the
bulk of the 3R phase. The position of the minimum in the first derivative obtained for the thinner samples, see inset in  Fig.~\ref{3R-thin}, cannot 
be determined with sufficient accuracy because it depends very much on the spatial position on the sample. Much better sample quality  of thin enough 3R phase samples is  necessary to check for  a semiconducting behavior with a larger energy gap than in the Bernal case as transport measurements suggest\cite{zor17}.
 Mean field theoretical studies indicate\cite{hya18} that the only possible origin for an 
 energy gap in the bulk of the 3R phase should be related to a
 spontaneous symmetry breaking, a feature that needs more work to verify its existence. 

\section{Conclusion}

Characteristic current-voltage curves obtained by tunneling atomic force spectroscopy (TUNA) on Bernal stacked graphite surfaces can successfully be reproduced by using classical self-consistent semiconductor modeling in conjunction with quantum mechanical tunneling current derivations, taking into account a non-zero band gap as well as reasonable values for the carrier concentration and effective masses of graphite. The best simulation  could be obtained for band gaps in
the range $( 12 \ldots 37)$~meV, in agreement with previous transport and Hall-effect  studies. The agreement between simulated and measured tunneling currents at 300~K should be considered as a demonstration that the assumption of graphite being a semiconductor with a narrow band gap
 is in line with the measured data. In particular our model explains the minimum in the $dI/dV$ curves at $V\simeq 120$~mV. It is the result of a delicate interplay of tunnel currents arising from valence- and conduction-band states at both positive and negative sample voltages. Furthermore, we showed that the lack of a voltage region without detectable current in the $I$-$V$ spectra, is not sufficient to exclude the presence of a band gap. In analogy, a voltage range without 
 detectable current is not one-to-one equal to the fundamental band gap.\cite{schnedler:2015a, portz:2017}   

In contrast to the results obtained from samples with Bernal stacking, the surface of samples with the 
rhombohedral stacking order indicate a clear maximum in the differential conductivity, in agreement with previous 
STS results obtained at much lower temperatures, suggesting the
existence of a flat band at the Fermi level. On the other hand, very thin samples with rhombohedral stacking order show a minimum
in the first derivative at positive voltages, whose voltage positions  depend very much on the sample location.


\section{Samples and Methods}
\label{sm}
\subsection{Samples}

The origin, thickness and method for electrical contacts of the measured samples are given in Table~\ref{samples}. 
The bulk HOPG Grade A sample, from which we prepared the samples HOPGA-Pd-1,  MG-Pd-11-1,  MG-Pd-11-2 and
A1 (see Table I), was obtained from Advanced Ceramics (now Momentive Performance Materials). The total impurity concentration (with exception of
H) of the HOPG and natural graphite samples is below 20~ppm, see Refs.~\cite{chap3,pre16} for more details.

\begin{table*}
  \caption{Characteristics of the  measured samples. The electrical contacts of the samples (last column) 
  were done through  lithography at the sample top surface or directly on the bottom sample surface using Pd-coated substrates. 3R means
  rhombohedral stacking order, 2H: Bernal stacking order.}
  \label{samples}
  \begin{tabular}{lllll}
    \hline
    Sample  & Thickness (nm)  & Origin & Stacking order & Contacts\\
    \hline
   NBF5-01-03   & 22 & Brazil mine & 3R &    lithography \\ 
   NBF5-01-05   & 40 & Brazil mine & mixture of 2H and 3R & lithography  \\
  NBF5-SNG-3  & 3 &  Brazil mine & 3R & lithography \\
   NBF5-01-09 &8 & Brazil mine & 2H & lithography\\
    HOPGA-Pd-1 & 300 & HOPG Grade A & 2H & Pd-coated\\
  MG-Pd-11-1 & 65 & HOPG Grade A & 2H & Pd-coated\\
    MG-Pd-11-2 & 40 & HOPG Grade A & 2H & Pd-coated\\
  A1 & 45 & HOPG Grade A & 2H & Pd-coated\\
    \hline
  \end{tabular}
\end{table*}

Graphite flakes were prepared using a mechanical cleavage. The method consists of mechanically gently rubbing a bulk graphite sample onto a thoroughly cleaned in ethanol substrate. After that, the substrate with the multigraphene flakes is well cleaned in an ultrasonic bath of highly concentrated acetone for one minute several times.

Different ways of electrical contacts fabrication were applied. The investigated graphite samples 
were placed on top of two differently prepared   silicon substrates with a  150~nm thick insulating silicon nitride (Si$_3$N$_4$) couting. 
On  one of the two  substrates we  patterned  electrodes on the samples top surface and on the insulating substrates 
using electron beam lithography followed by sputtering of a bilayer Cr/Au (5~nm/30~nm). Other graphite samples were electrically 
contacted  by depositing their bottom surface to a  100~nm thick Pd layer  sputtered on the Si$_3$N$_4$ coating (after depositing  a buffer layer of 31~nm thick Cr).

\begin{figure} [h] 
 \includegraphics[width=1.05\columnwidth]{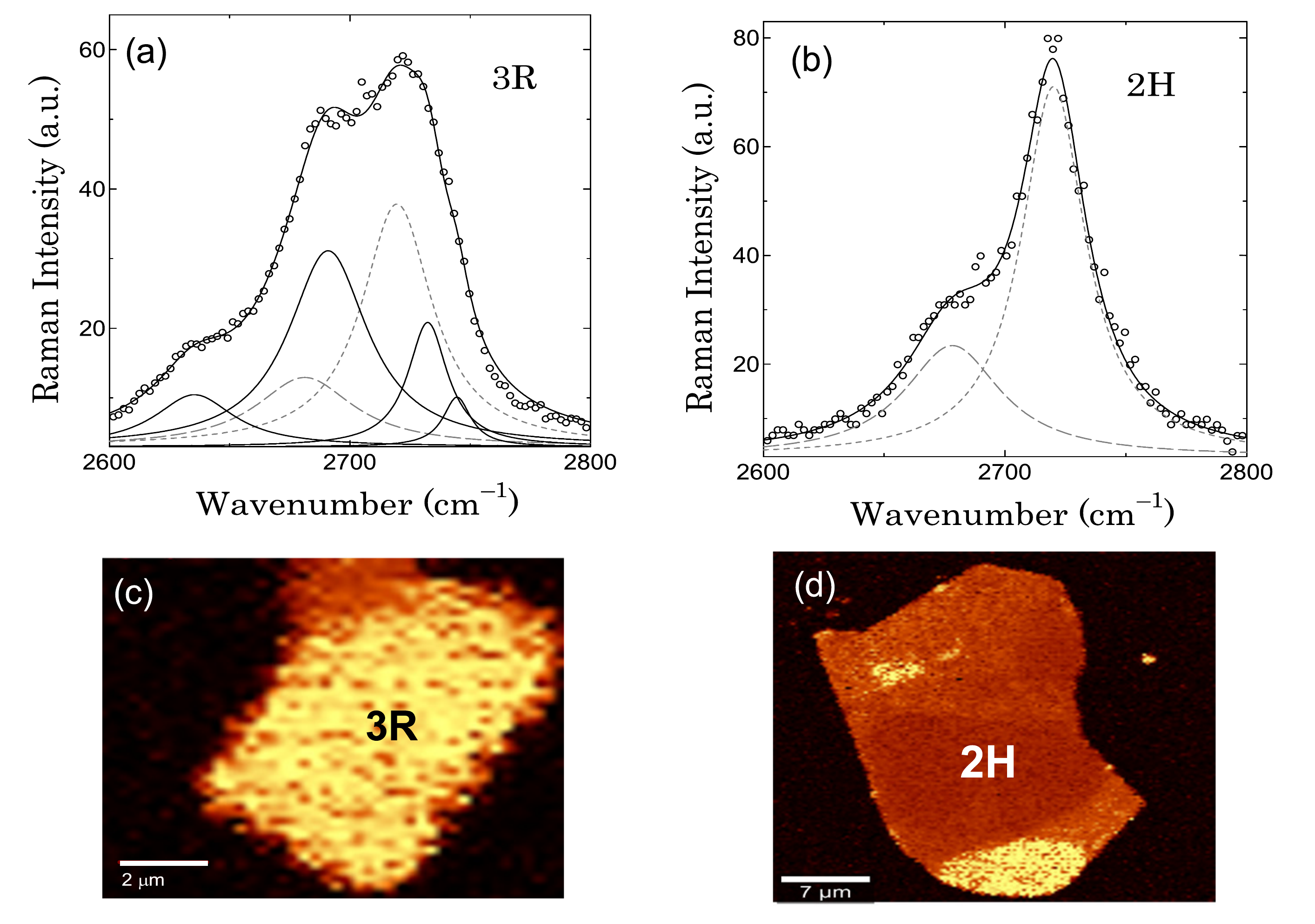}  
  \caption{ (a) Average of the 2D band of the rhombohedral phase (3R) showing the six processes due to ABC stacking of the NBF5-01-03 natural graphite flake sample.  (b) Average of the  2D band of the Bernal phase (2H) showing the two process due to ABA stacking of MGPd-11 HOPG flake sample. (c) Raman image of the spatial distribution of the 2D band width. The yellow color area shows the 3R phase.(d) Raman image for the spatial distribution of the 2D band. In this case the red color area represents the Bernal phase (2H).}
  \label{raman}
\end{figure}

\subsection{Raman characterization}
Microscopic analysis of the graphite stacking orders and sample quality were realized  by Raman spectroscopy measurements. Raman spectra of multilayer graphene samples were obtained with a confocal micro-Raman microscope WITec alpha 300+ at 532~nm wavelength (green) at ambient temperature, see Fig.~\ref{raman} as example. All Raman measurements were performed using a grating of 1800 grooves mm$^{-1}$, 100x objectives (NA 0.9) and incident laser power of 1~mW. Lateral resolution of the confocal micro-Raman microscope  was  $\sim 300~$nm and the depth resolution $\sim 800~$nm.

The Raman results 
revealed Bernal and rhombohedral stacking orders as well as a mixture of those phases depending on sample. 
The areas with different stacking orders have been localized by surface scanning and registration of spectra at each point in increments of 100 nm  with the confocal micro-Raman microscope. Afterwards, the function of automatic recognition of the given shapes of the spectrum curves showed the regions with different stacking.

Figures~\ref{raman}(a,b) present the Raman spectra in the 2D Raman bands  for rhombohedral (3R) area of the NBFG5-01 sample and Bernal (2H) area of the MGPd-11 sample. The spectra  are the average obtained  by  hyperspectral scanning within the respective regions.  
Figures~\ref{raman}(c,d) show the Raman image that is obtained analyzing the spatial distribution of the 2D band width of the NBF5-01 and MGPd-11 samples.
Following recent publications \cite{con11,hen16,tor17,ram20} 
there are three main Raman scattering features in rhombohedral stacking order of the graphite structure. The main one that can be used to
find the 3R phase is the absorption at the G'-band with a broad peak around $\simeq 2700~$cm$^{-1}$, as shown in Fig.~\ref{raman}(a), 
in clear contrast to
the corresponding peak measured in the 2H phase, see Fig.~\ref{raman}(b). 
According to the Raman results the investigated samples presented in this study are of 
high structural quality  due to the absence of the disorder-related D-peak around $\simeq 1350~$cm$^{-1}$ (not shown).

 These results indicate that while the smaller area in the NBF5-01 sample has a Bernal stacking order, the larger area consists of rhombohedral one. The opposite is observed in the other sample shown in Fig.~\ref{raman}(d).
\begin{figure}[h]
 \includegraphics[width=1\columnwidth]{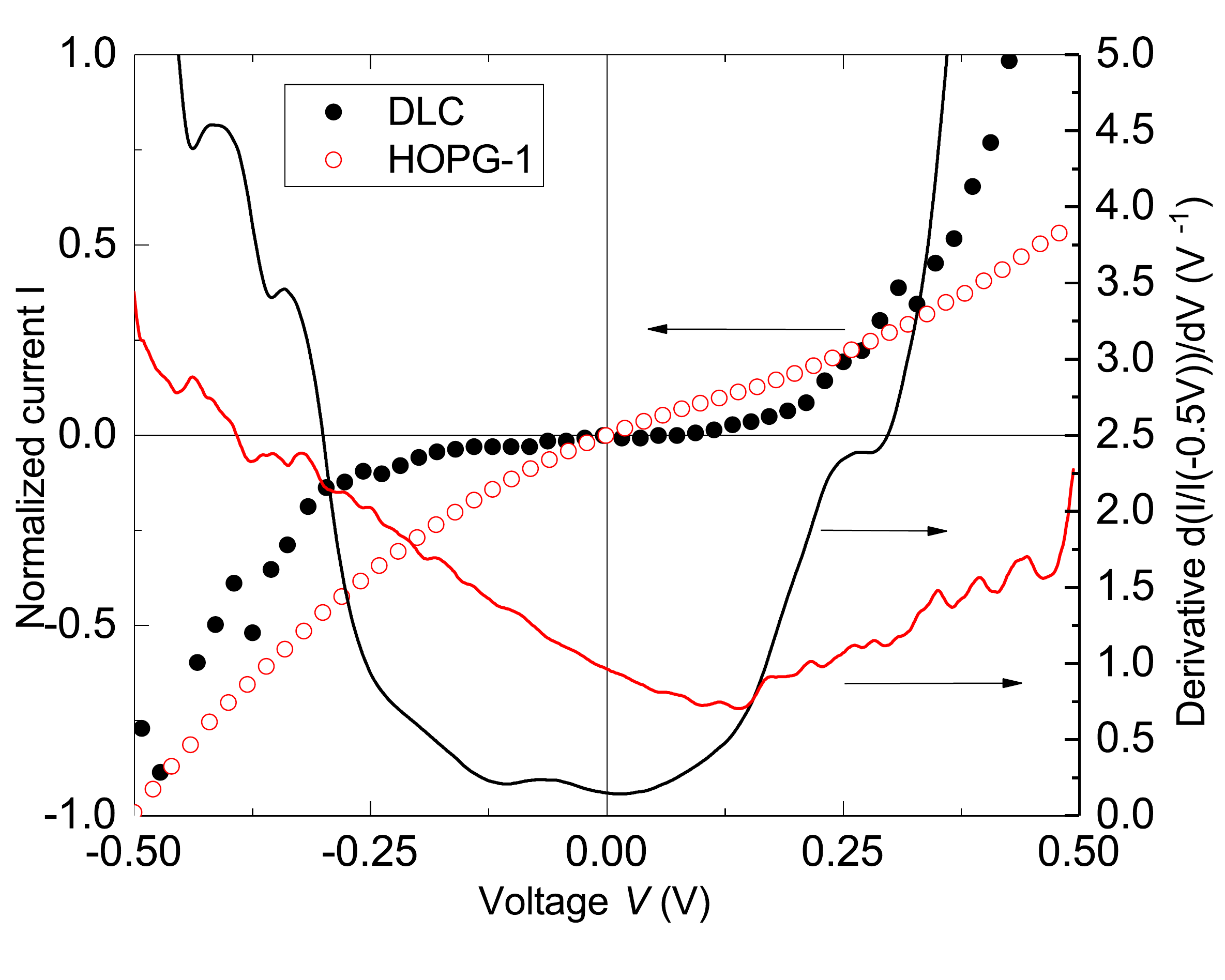}  
  \caption{TUNA results of the current-voltage characteristic curves (left y-axis)  and their first derivatives (right y-axis) 
  measured on a diamond-like carbon film (DLC)  and on the  position 1 of
  the bulk graphite surface of Fig.~\ref{gra}(a). 
  The maximum current values used for normalization were: 59~pA $(\bullet)$ and 217~pA ($\circ)$. The red line of the first derivative
  corresponds to the graphite sample and the black one to the DLC film. The derivatives were obtained from the normalized current data
  at -0.5~V. }
  \label{dlc}
\end{figure}

\subsection{TUNA measurements}

 Local current-voltage ($I$-$V$) curves measurements were performed with a Bruker Dimension Icon Scanning Probe Microscope equipped with PeakForce Tunneling AFM module (PF-TUNA)\cite{PF}. The PeakForce Tapping mode is  based on a quick contact interaction of the probe and the sample (tens to hundreds of microseconds). PeakForce TUNA mode with a  bandwidth of 15~kHz allows the measurement of a  current  averaged  over the full tapping cycle. All measurements were conducted at room temperature and at ambient conditions.  
 Pt-Ir-coated silicon nitride probes with a nominal radius of 25~nm (PF-TUNA, spring constant = 0.4 N/m, resonant frequency = 70 kHz) have been used. PeakForce Tapping Amplitude was set to 150~nm. A current sensitivity of 100 pA/V was used in all measurements. 
The bias voltage was swept  from -500~mV to 500~mV on flat surface regions. 
 In order to decrease the noise in the data, the $I$-$V$ curves shown in this study were taken from an average of 25 consecutive $I$-$V$ ramps at a single point.

The measurement and calibration of the $I-V$ curves at certain parts of the selected samples were performed as follows: First the $I-V$ curves were measured on a test sample - a floppy disk. After checking the operability, the cantilever was moved to the graphite sample and a $ 20 \time 20~$nm$^2$ area was scanned in order to localize the flat areas without irregularities. Then a  point was chosen in an area of $ 5 \times 5~$nm$^2$ or  $ 1 \times 1~$nm$^2$ to register the I-V curves. The PeakForce setpoint (trigger force) was chosen as small as possible in order not to damage the surface.

To rule out possible artifacts (e.g., when a piece of graphite sticks at the tip), to check the reproducibility of the measurements and 
the state of the used Pt-Ir probes, three different tests were used, namely:  After each measurement the response of the Pt-Ir probe was examined using the reference sample (FD sample, 12~mm from Bruker). Furthermore,
scanning electron microscope (SEM) was used to  check for the state of the probes tips. In order to avoid distortion in the data, each probe was used not more than 25~h. 
After that  we continued with  a new probe. In the case of the samples with the 3R phase  two different probes were used to check for the reproducibility of  
the unusual  behavior at low voltages.

Finally, to verify that the shift of the minima in $dI/dV$ to positive bias voltages obtained in the graphite samples is 
not an artifact, we selected a $n$-type semiconducting sample with larger band gap than graphite, but still relatively narrow. 
Just after measuring one sample with Bernal stacking and without changing the probe or any sets of the microscope, we  
measured a hydrogenated diamond like carbon (DLC) film, the one usually used in old magnetic hard disks. Upon
 hydrogenation and defect concentration the energy gap of the DLC films can be as low as a few hundreds of meV. 
 The measured $I$-$V$ curve of the DLC film is shown in Fig.~\ref{dlc} together
with its differential conductance (right axis). For comparison,
in the same figure   we include one of the curves obtained from the bulk HOPG sample
shown in Fig.~\ref{gra}(a).  The $I$-$V$ curve  of the DLC film and its derivative
indicate an energy gap $E_g \gtrsim 0.25$~eV. 
Note that the $I$-$V$ curve of the DLC film shows an opposite asymmetry as for graphite, i.e. it is shifted to 
negative bias voltages. 

\begin{figure}[h]
 \includegraphics[width=1\columnwidth]{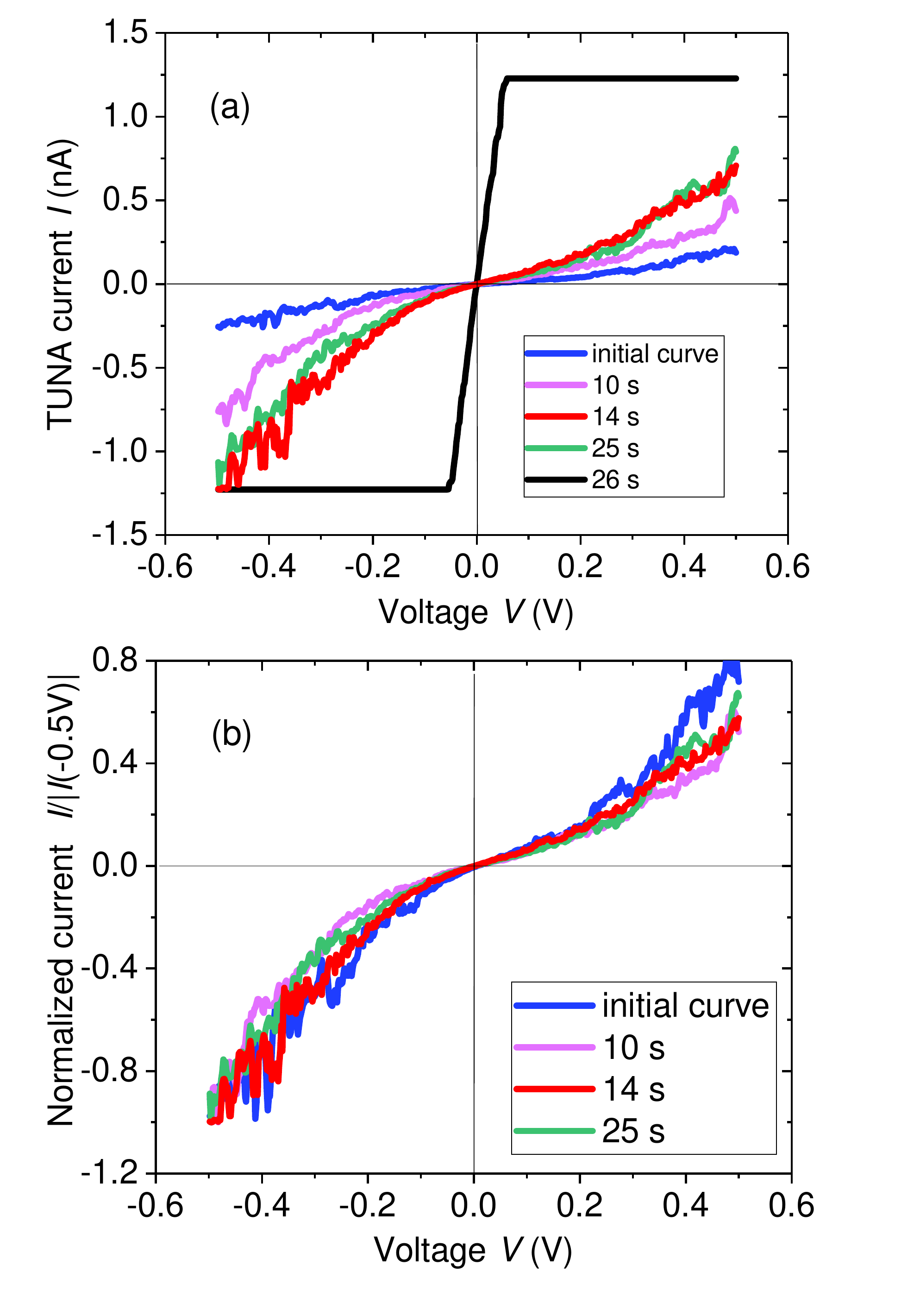}  
  \caption{(a) Single (no averaging) TUNA current vs. bias voltage curves obtained 
  on sample HOPGA-Pd-1 (2H phase) at a fixed location in Continuous Ramping mode 
  at different times. After 26~s the tunneling current saturates 
  already at low bias voltages just before a direct contact between tip and the sample surface occurs. 
  (b) The same data as in (a) in the tunneling regime but
  normalized by their corresponding  currents at -0.5~V.}
  \label{time}
\end{figure}

The  TUNA equipment allows in principle a constant current scan to get  a scan image of the 
conductance variation across the sample surface. This useful tool cannot be easily used for graphite due to 
the soft $c-$axis elastic module  of the graphite structure. When one uses this mode and due to the soft elastic module a
 time dependence  in the $I$-$V$ curves is observed. 
Measurements of the time-dependence of the $I$-$V$ spectra were carried out with the graphite sample HOPG-Pd-1  
using the spectroscopy mode in the PeakForce TUNA mode. 
In order to obtain a series of $I$-$V$ curves at one fixed position the Continuous Ramping mode 
was used with the parameters: 2~V set point, 100pA/V sensitivity and ramp rate of 1.03 Hz. 
The cycle was repeated automatically after certain time. Overall, 26 $I$-$V$ curves were obtained within 26~s, some of them
are shown in Fig.~\ref{time}. These measurements indicate a variation of the tunneling current with time. The observed behavior
is  similar to the one observed from tunneling regime to point-contact changing the tip-sample distance in Ref.~\cite{agr92}. 
In the tunneling regime, i.e. at low enough current amplitudes and 
before a direct contact is achieved, all $I$-$V$ curves 
show the same behavior after  normalization, see Fig.~\ref{time}(b). 
This effect should be taken into account when measuring electrical characteristics of soft 
materials with PF-TUNA mode.
From the technical point of view, the obtained results indicate that the  TUNA  technique
allows  spectroscopy studies in  natural and pyrolytic 
 graphite samples. Care must be taken with time dependent effects due to the soft 
 $c-$axis module of graphite and the set-point selected in the TUNA electronics.\\

{\em Author Contributions}: R.A. carried out the TUNA measurements. 
R.A. and A.C.  carried out the sample preparation. T.V. and I.E-L. performed
the Raman measurements. A.C. and R.A. contributed to the interpretation of the Raman results. M.Sc. contributed to the interpretation of the results and 
performed all numerical studies. W.H., R.E.D-B.  and P.E. contributed to the interpretation of the results. 
M.St. helped in the measurements and purchasing the microscope with all the used options. 
P.D.E.  conceived and planned the experiments. P.D.E. and M.Sc.  took the lead in writing the manuscript.
 All authors provided critical feedback and helped shape the research, analysis, and manuscript.\\

The authors declare no competing financial interest.\\
 
Acknowledgement:
The author P.D.E.  thanks Tero Heikkil\"a, Francesco Mauri, Vladimir Volovik, Jose-Gabriel Rodrigo,
 Andrei Pavlov and Eberhard K. U. Gross for fruitful discussions. We thank Laetitia Bettmann for her help
 during the studies. 
This work has been possible with the support of the Saechsischen Aufbaubank (SAB) and the European Regional
Development Fund, Grant Nr.: 231301388.
A.C. acknowledges  2MI and VRS Eireli Research \& Development for the financial support and Nacional do Grafite, MG (Brasil), for the samples.


\begin{thebibliography}{77}%
\makeatletter
\providecommand \@ifxundefined [1]{%
 \@ifx{#1\undefined}
}%
\providecommand \@ifnum [1]{%
 \ifnum #1\expandafter \@firstoftwo
 \else \expandafter \@secondoftwo
 \fi
}%
\providecommand \@ifx [1]{%
 \ifx #1\expandafter \@firstoftwo
 \else \expandafter \@secondoftwo
 \fi
}%
\providecommand \natexlab [1]{#1}%
\providecommand \enquote  [1]{``#1''}%
\providecommand \bibnamefont  [1]{#1}%
\providecommand \bibfnamefont [1]{#1}%
\providecommand \citenamefont [1]{#1}%
\providecommand \href@noop [0]{\@secondoftwo}%
\providecommand \href [0]{\begingroup \@sanitize@url \@href}%
\providecommand \@href[1]{\@@startlink{#1}\@@href}%
\providecommand \@@href[1]{\endgroup#1\@@endlink}%
\providecommand \@sanitize@url [0]{\catcode `\\12\catcode `\$12\catcode
  `\&12\catcode `\#12\catcode `\^12\catcode `\_12\catcode `\%12\relax}%
\providecommand \@@startlink[1]{}%
\providecommand \@@endlink[0]{}%
\providecommand \url  [0]{\begingroup\@sanitize@url \@url }%
\providecommand \@url [1]{\endgroup\@href {#1}{\urlprefix }}%
\providecommand \urlprefix  [0]{URL }%
\providecommand \Eprint [0]{\href }%
\providecommand \doibase [0]{http://dx.doi.org/}%
\providecommand \selectlanguage [0]{\@gobble}%
\providecommand \bibinfo  [0]{\@secondoftwo}%
\providecommand \bibfield  [0]{\@secondoftwo}%
\providecommand \translation [1]{[#1]}%
\providecommand \BibitemOpen [0]{}%
\providecommand \bibitemStop [0]{}%
\providecommand \bibitemNoStop [0]{.\EOS\space}%
\providecommand \EOS [0]{\spacefactor3000\relax}%
\providecommand \BibitemShut  [1]{\csname bibitem#1\endcsname}%
\let\auto@bib@innerbib\@empty
\bibitem [{\citenamefont {Zhang}\ \emph {et~al.}(2010)\citenamefont {Zhang},
  \citenamefont {Min}, \citenamefont {Polini},\ and\ \citenamefont
  {MacDonald}}]{zha10}%
  \BibitemOpen
  \bibfield  {author} {\bibinfo {author} {\bibfnamefont {F.}~\bibnamefont
  {Zhang}}, \bibinfo {author} {\bibfnamefont {H.}~\bibnamefont {Min}}, \bibinfo
  {author} {\bibfnamefont {M.}~\bibnamefont {Polini}}, \ and\ \bibinfo {author}
  {\bibfnamefont {A.~H.}\ \bibnamefont {MacDonald}},\ }\href {\doibase
  10.1103/PhysRevB.81.041402} {\bibfield  {journal} {\bibinfo  {journal} {Phys.
  Rev. B}\ }\textbf {\bibinfo {volume} {81}},\ \bibinfo {pages} {041402}
  (\bibinfo {year} {2010})}\BibitemShut {NoStop}%
\bibitem [{\citenamefont {Garc\'ia}\ \emph {et~al.}(2012)\citenamefont
  {Garc\'ia}, \citenamefont {Esquinazi}, \citenamefont {Barzola-Quiquia},\ and\
  \citenamefont {Dusari}}]{gar12}%
  \BibitemOpen
  \bibfield  {author} {\bibinfo {author} {\bibfnamefont {N.}~\bibnamefont
  {Garc\'ia}}, \bibinfo {author} {\bibfnamefont {P.}~\bibnamefont {Esquinazi}},
  \bibinfo {author} {\bibfnamefont {J.}~\bibnamefont {Barzola-Quiquia}}, \ and\
  \bibinfo {author} {\bibfnamefont {S.}~\bibnamefont {Dusari}},\ }\href@noop {}
  {\bibfield  {journal} {\bibinfo  {journal} {New Journal of Physics}\ }\textbf
  {\bibinfo {volume} {14}},\ \bibinfo {pages} {053015} (\bibinfo {year}
  {2012})}\BibitemShut {NoStop}%
\bibitem [{\citenamefont {Kelly}(1981)}]{kelly}%
  \BibitemOpen
  \bibfield  {author} {\bibinfo {author} {\bibfnamefont {B.~T.}\ \bibnamefont
  {Kelly}},\ }\href@noop {} {\emph {\bibinfo {title} {Physics of Graphite}}}\
  (\bibinfo  {publisher} {London: Applied Science Publishers},\ \bibinfo {year}
  {1981})\BibitemShut {NoStop}%
\bibitem [{\citenamefont {Lin}\ \emph {et~al.}(2012)\citenamefont {Lin},
  \citenamefont {Li}, \citenamefont {Liu}, \citenamefont {Song}, \citenamefont
  {He}, \citenamefont {Hu}, \citenamefont {Guo},\ and\ \citenamefont
  {Ye}}]{lin12}%
  \BibitemOpen
  \bibfield  {author} {\bibinfo {author} {\bibfnamefont {Q.}~\bibnamefont
  {Lin}}, \bibinfo {author} {\bibfnamefont {T.}~\bibnamefont {Li}}, \bibinfo
  {author} {\bibfnamefont {Z.}~\bibnamefont {Liu}}, \bibinfo {author}
  {\bibfnamefont {Y.}~\bibnamefont {Song}}, \bibinfo {author} {\bibfnamefont
  {L.}~\bibnamefont {He}}, \bibinfo {author} {\bibfnamefont {Z.}~\bibnamefont
  {Hu}}, \bibinfo {author} {\bibfnamefont {Q.}~\bibnamefont {Guo}}, \ and\
  \bibinfo {author} {\bibfnamefont {H.}~\bibnamefont {Ye}},\ }\href@noop {}
  {\bibfield  {journal} {\bibinfo  {journal} {Carbon}\ }\textbf {\bibinfo
  {volume} {50}},\ \bibinfo {pages} {2369} (\bibinfo {year}
  {2012})}\BibitemShut {NoStop}%
\bibitem [{\citenamefont {Precker}\ \emph {et~al.}(2016)\citenamefont
  {Precker}, \citenamefont {Esquinazi}, \citenamefont {Champi}, \citenamefont
  {Barzola-Quiquia}, \citenamefont {Zoraghi}, \citenamefont
  {Mui{\~n}os-Landin}, \citenamefont {Setzer}, \citenamefont {B{\"o}hlmann},
  \citenamefont {Spemann}, \citenamefont {Meijer}, \citenamefont {Muenster},
  \citenamefont {Baehre}, \citenamefont {Kloess},\ and\ \citenamefont
  {Beth}}]{pre16}%
  \BibitemOpen
  \bibfield  {author} {\bibinfo {author} {\bibfnamefont {C.~E.}\ \bibnamefont
  {Precker}}, \bibinfo {author} {\bibfnamefont {P.~D.}\ \bibnamefont
  {Esquinazi}}, \bibinfo {author} {\bibfnamefont {A.}~\bibnamefont {Champi}},
  \bibinfo {author} {\bibfnamefont {J.}~\bibnamefont {Barzola-Quiquia}},
  \bibinfo {author} {\bibfnamefont {M.}~\bibnamefont {Zoraghi}}, \bibinfo
  {author} {\bibfnamefont {S.}~\bibnamefont {Mui{\~n}os-Landin}}, \bibinfo
  {author} {\bibfnamefont {A.}~\bibnamefont {Setzer}}, \bibinfo {author}
  {\bibfnamefont {W.}~\bibnamefont {B{\"o}hlmann}}, \bibinfo {author}
  {\bibfnamefont {D.}~\bibnamefont {Spemann}}, \bibinfo {author} {\bibfnamefont
  {J.}~\bibnamefont {Meijer}}, \bibinfo {author} {\bibfnamefont
  {T.}~\bibnamefont {Muenster}}, \bibinfo {author} {\bibfnamefont
  {O.}~\bibnamefont {Baehre}}, \bibinfo {author} {\bibfnamefont
  {G.}~\bibnamefont {Kloess}}, \ and\ \bibinfo {author} {\bibfnamefont
  {H.}~\bibnamefont {Beth}},\ }\href {\doibase
  https://doi.org/10.1088/1367-2630/18/11/113041} {\bibfield  {journal}
  {\bibinfo  {journal} {New J. Phys.}\ }\textbf {\bibinfo {volume} {18}},\
  \bibinfo {pages} {113041} (\bibinfo {year} {2016})}\BibitemShut {NoStop}%
\bibitem [{\citenamefont {Barzola-Quiquia}\ \emph {et~al.}(2008)\citenamefont
  {Barzola-Quiquia}, \citenamefont {Yao}, \citenamefont {R\"odiger},
  \citenamefont {Schindler},\ and\ \citenamefont {Esquinazi}}]{bar08}%
  \BibitemOpen
  \bibfield  {author} {\bibinfo {author} {\bibfnamefont {J.}~\bibnamefont
  {Barzola-Quiquia}}, \bibinfo {author} {\bibfnamefont {J.-L.}\ \bibnamefont
  {Yao}}, \bibinfo {author} {\bibfnamefont {P.}~\bibnamefont {R\"odiger}},
  \bibinfo {author} {\bibfnamefont {K.}~\bibnamefont {Schindler}}, \ and\
  \bibinfo {author} {\bibfnamefont {P.}~\bibnamefont {Esquinazi}},\ }\href@noop
  {} {\bibfield  {journal} {\bibinfo  {journal} {phys. stat. sol. (a)}\
  }\textbf {\bibinfo {volume} {205}},\ \bibinfo {pages} {2924} (\bibinfo {year}
  {2008})}\BibitemShut {NoStop}%
\bibitem [{\citenamefont {Ballestar}\ \emph {et~al.}(2013)\citenamefont
  {Ballestar}, \citenamefont {Barzola-Quiquia}, \citenamefont {Scheike},\ and\
  \citenamefont {Esquinazi}}]{bal13}%
  \BibitemOpen
  \bibfield  {author} {\bibinfo {author} {\bibfnamefont {A.}~\bibnamefont
  {Ballestar}}, \bibinfo {author} {\bibfnamefont {J.}~\bibnamefont
  {Barzola-Quiquia}}, \bibinfo {author} {\bibfnamefont {T.}~\bibnamefont
  {Scheike}}, \ and\ \bibinfo {author} {\bibfnamefont {P.}~\bibnamefont
  {Esquinazi}},\ }\href@noop {} {\bibfield  {journal} {\bibinfo  {journal} {New
  J. Phys.}\ }\textbf {\bibinfo {volume} {15}},\ \bibinfo {pages} {023024}
  (\bibinfo {year} {2013})}\BibitemShut {NoStop}%
\bibitem [{\citenamefont {Scheike}\ \emph {et~al.}(2013)\citenamefont
  {Scheike}, \citenamefont {Esquinazi}, \citenamefont {Setzer},\ and\
  \citenamefont {B\"ohlmann}}]{schcar}%
  \BibitemOpen
  \bibfield  {author} {\bibinfo {author} {\bibfnamefont {T.}~\bibnamefont
  {Scheike}}, \bibinfo {author} {\bibfnamefont {P.}~\bibnamefont {Esquinazi}},
  \bibinfo {author} {\bibfnamefont {A.}~\bibnamefont {Setzer}}, \ and\ \bibinfo
  {author} {\bibfnamefont {W.}~\bibnamefont {B\"ohlmann}},\ }\href@noop {}
  {\bibfield  {journal} {\bibinfo  {journal} {Carbon}\ }\textbf {\bibinfo
  {volume} {59}},\ \bibinfo {pages} {140} (\bibinfo {year} {2013})}\BibitemShut
  {NoStop}%
\bibitem [{\citenamefont {Ballestar}\ \emph {et~al.}(2015)\citenamefont
  {Ballestar}, \citenamefont {Esquinazi},\ and\ \citenamefont
  {B\"ohlmann}}]{bal15}%
  \BibitemOpen
  \bibfield  {author} {\bibinfo {author} {\bibfnamefont {A.}~\bibnamefont
  {Ballestar}}, \bibinfo {author} {\bibfnamefont {P.}~\bibnamefont
  {Esquinazi}}, \ and\ \bibinfo {author} {\bibfnamefont {W.}~\bibnamefont
  {B\"ohlmann}},\ }\href@noop {} {\bibfield  {journal} {\bibinfo  {journal}
  {Phys. Rev. B}\ }\textbf {\bibinfo {volume} {91}},\ \bibinfo {pages} {014502}
  (\bibinfo {year} {2015})}\BibitemShut {NoStop}%
\bibitem [{\citenamefont {Esquinazi}\ and\ \citenamefont
  {Lysogorskiy}(2016)}]{chap7}%
  \BibitemOpen
  \bibfield  {author} {\bibinfo {author} {\bibfnamefont {P.~D.}\ \bibnamefont
  {Esquinazi}}\ and\ \bibinfo {author} {\bibfnamefont {Y.}~\bibnamefont
  {Lysogorskiy}},\ }\enquote {\bibinfo {title} {Experimental evidence for the
  existence of interfaces in graphite and their relation to the observed
  metallic and superconducting behavior},}\ \ (\bibinfo  {publisher} {P.
  Esquinazi (ed.), Springer International Publishing AG Switzerland},\ \bibinfo
  {year} {2016})\ Chap.~\bibinfo {chapter} {7}, pp.\ \bibinfo {pages}
  {145--179}\BibitemShut {NoStop}%
\bibitem [{\citenamefont {Zoraghi}\ \emph {et~al.}(2018)\citenamefont
  {Zoraghi}, \citenamefont {Barzola-Quiquia}, \citenamefont {Stiller},
  \citenamefont {Esquinazi},\ and\ \citenamefont {Estrela-Lopis}}]{zor18}%
  \BibitemOpen
  \bibfield  {author} {\bibinfo {author} {\bibfnamefont {M.}~\bibnamefont
  {Zoraghi}}, \bibinfo {author} {\bibfnamefont {J.}~\bibnamefont
  {Barzola-Quiquia}}, \bibinfo {author} {\bibfnamefont {M.}~\bibnamefont
  {Stiller}}, \bibinfo {author} {\bibfnamefont {P.~D.}\ \bibnamefont
  {Esquinazi}}, \ and\ \bibinfo {author} {\bibfnamefont {I.}~\bibnamefont
  {Estrela-Lopis}},\ }\href {\doibase
  https://doi.org/10.1016/j.carbon.2018.07.070} {\bibfield  {journal} {\bibinfo
   {journal} {Carbon}\ }\textbf {\bibinfo {volume} {139}},\ \bibinfo {pages}
  {1074 } (\bibinfo {year} {2018})}\BibitemShut {NoStop}%
\bibitem [{\citenamefont {Precker}\ \emph {et~al.}(2019)\citenamefont
  {Precker}, \citenamefont {Barzola-Quiquia}, \citenamefont {Esquinazi},
  \citenamefont {Stiller}, \citenamefont {Chan}, \citenamefont {Jaime},
  \citenamefont {Zhang},\ and\ \citenamefont {Grundmann}}]{pre19}%
  \BibitemOpen
  \bibfield  {author} {\bibinfo {author} {\bibfnamefont {C.~E.}\ \bibnamefont
  {Precker}}, \bibinfo {author} {\bibfnamefont {J.}~\bibnamefont
  {Barzola-Quiquia}}, \bibinfo {author} {\bibfnamefont {P.~D.}\ \bibnamefont
  {Esquinazi}}, \bibinfo {author} {\bibfnamefont {M.}~\bibnamefont {Stiller}},
  \bibinfo {author} {\bibfnamefont {M.~K.}\ \bibnamefont {Chan}}, \bibinfo
  {author} {\bibfnamefont {M.}~\bibnamefont {Jaime}}, \bibinfo {author}
  {\bibfnamefont {Z.}~\bibnamefont {Zhang}}, \ and\ \bibinfo {author}
  {\bibfnamefont {M.}~\bibnamefont {Grundmann}},\ }\href {\doibase
  10.1002/adem.201970039} {\bibfield  {journal} {\bibinfo  {journal} {Advanced
  Engineering Materials}\ }\textbf {\bibinfo {volume} {21}},\ \bibinfo {pages}
  {1970039} (\bibinfo {year} {2019})}\BibitemShut {NoStop}%
\bibitem [{\citenamefont {Kuwabara}\ \emph {et~al.}(1990)\citenamefont
  {Kuwabara}, \citenamefont {Clarke},\ and\ \citenamefont {Smith}}]{kuw90}%
  \BibitemOpen
  \bibfield  {author} {\bibinfo {author} {\bibfnamefont {M.}~\bibnamefont
  {Kuwabara}}, \bibinfo {author} {\bibfnamefont {D.~R.}\ \bibnamefont
  {Clarke}}, \ and\ \bibinfo {author} {\bibfnamefont {A.~A.}\ \bibnamefont
  {Smith}},\ }\href@noop {} {\bibfield  {journal} {\bibinfo  {journal} {Appl.
  Phys. Lett.}\ }\textbf {\bibinfo {volume} {56}},\ \bibinfo {pages} {2396}
  (\bibinfo {year} {1990})}\BibitemShut {NoStop}%
\bibitem [{\citenamefont {Flores}\ \emph {et~al.}(2013)\citenamefont {Flores},
  \citenamefont {Cisternas}, \citenamefont {Correa},\ and\ \citenamefont
  {Vargas}}]{flo13}%
  \BibitemOpen
  \bibfield  {author} {\bibinfo {author} {\bibfnamefont {M.}~\bibnamefont
  {Flores}}, \bibinfo {author} {\bibfnamefont {E.}~\bibnamefont {Cisternas}},
  \bibinfo {author} {\bibfnamefont {J.}~\bibnamefont {Correa}}, \ and\ \bibinfo
  {author} {\bibfnamefont {P.}~\bibnamefont {Vargas}},\ }\href@noop {}
  {\bibfield  {journal} {\bibinfo  {journal} {Chemical Physics}\ }\textbf
  {\bibinfo {volume} {423}},\ \bibinfo {pages} {49} (\bibinfo {year}
  {2013})}\BibitemShut {NoStop}%
\bibitem [{\citenamefont {Miller}\ \emph {et~al.}(2010)\citenamefont {Miller},
  \citenamefont {Kubista}, \citenamefont {Rutter}, \citenamefont {Ruan},
  \citenamefont {de~Heer}, \citenamefont {First},\ and\ \citenamefont
  {Stroscio}}]{mil10}%
  \BibitemOpen
  \bibfield  {author} {\bibinfo {author} {\bibfnamefont {D.~L.}\ \bibnamefont
  {Miller}}, \bibinfo {author} {\bibfnamefont {K.~D.}\ \bibnamefont {Kubista}},
  \bibinfo {author} {\bibfnamefont {G.~M.}\ \bibnamefont {Rutter}}, \bibinfo
  {author} {\bibfnamefont {M.}~\bibnamefont {Ruan}}, \bibinfo {author}
  {\bibfnamefont {W.~A.}\ \bibnamefont {de~Heer}}, \bibinfo {author}
  {\bibfnamefont {P.~N.}\ \bibnamefont {First}}, \ and\ \bibinfo {author}
  {\bibfnamefont {J.~A.}\ \bibnamefont {Stroscio}},\ }\href@noop {} {\bibfield
  {journal} {\bibinfo  {journal} {Phys. Rev. B}\ }\textbf {\bibinfo {volume}
  {81}},\ \bibinfo {pages} {125427} (\bibinfo {year} {2010})}\BibitemShut
  {NoStop}%
\bibitem [{\citenamefont {Brihuega}\ \emph {et~al.}(2012)\citenamefont
  {Brihuega}, \citenamefont {Mallet}, \citenamefont {Gonz\'alez-Herrero},
  \citenamefont {de~Laissardi\`ere}, \citenamefont {Ugeda}, \citenamefont
  {Magaud}, \citenamefont {G\'omez-Rodr\'iguez}, \citenamefont {Yndur\'ain},\
  and\ \citenamefont {Veuillen}}]{bri12}%
  \BibitemOpen
  \bibfield  {author} {\bibinfo {author} {\bibfnamefont {I.}~\bibnamefont
  {Brihuega}}, \bibinfo {author} {\bibfnamefont {P.}~\bibnamefont {Mallet}},
  \bibinfo {author} {\bibfnamefont {H.}~\bibnamefont {Gonz\'alez-Herrero}},
  \bibinfo {author} {\bibfnamefont {G.~T.}\ \bibnamefont {de~Laissardi\`ere}},
  \bibinfo {author} {\bibfnamefont {M.~M.}\ \bibnamefont {Ugeda}}, \bibinfo
  {author} {\bibfnamefont {L.}~\bibnamefont {Magaud}}, \bibinfo {author}
  {\bibfnamefont {J.~M.}\ \bibnamefont {G\'omez-Rodr\'iguez}}, \bibinfo
  {author} {\bibfnamefont {F.}~\bibnamefont {Yndur\'ain}}, \ and\ \bibinfo
  {author} {\bibfnamefont {J.-Y.}\ \bibnamefont {Veuillen}},\ }\href@noop {}
  {\bibfield  {journal} {\bibinfo  {journal} {Phys. Rev. Lett.}\ }\textbf
  {\bibinfo {volume} {109}},\ \bibinfo {pages} {196802} (\bibinfo {year}
  {2012})}\BibitemShut {NoStop}%
\bibitem [{\citenamefont {Lu}\ \emph {et~al.}(2006)\citenamefont {Lu},
  \citenamefont {Mu{\~n}oz}, \citenamefont {Steplecaru}, \citenamefont {Hao},
  \citenamefont {Bai}, \citenamefont {Garc{\'\i}a}, \citenamefont {Schindler},\
  and\ \citenamefont {Esquinazi}}]{lu06}%
  \BibitemOpen
  \bibfield  {author} {\bibinfo {author} {\bibfnamefont {Y.}~\bibnamefont
  {Lu}}, \bibinfo {author} {\bibfnamefont {M.}~\bibnamefont {Mu{\~n}oz}},
  \bibinfo {author} {\bibfnamefont {C.~S.}\ \bibnamefont {Steplecaru}},
  \bibinfo {author} {\bibfnamefont {C.}~\bibnamefont {Hao}}, \bibinfo {author}
  {\bibfnamefont {M.}~\bibnamefont {Bai}}, \bibinfo {author} {\bibfnamefont
  {N.}~\bibnamefont {Garc{\'\i}a}}, \bibinfo {author} {\bibfnamefont
  {K.}~\bibnamefont {Schindler}}, \ and\ \bibinfo {author} {\bibfnamefont
  {P.}~\bibnamefont {Esquinazi}},\ }\href@noop {} {\bibfield  {journal}
  {\bibinfo  {journal} {Phys. Rev. Lett.}\ }\textbf {\bibinfo {volume} {97}},\
  \bibinfo {pages} {076805} (\bibinfo {year} {2006})},\ \bibinfo {note} {see
  also the comment by S. Sadewasser and Th. Glatzel, Phys. Rev. Lett. {\bf 98},
  269701 (2007) and the reply by Lu et al., {\it idem} {\bf 98}, 269702 (2007)
  and also R. Proksch, Appl. Phys. Lett. {\bf 89}, 113121 (2006).}\BibitemShut
  {Stop}%
\bibitem [{\citenamefont {McClure}(1956)}]{mcclure56}%
  \BibitemOpen
  \bibfield  {author} {\bibinfo {author} {\bibfnamefont {J.~W.}\ \bibnamefont
  {McClure}},\ }\href@noop {} {\bibfield  {journal} {\bibinfo  {journal} {Phys.
  Rev.}\ }\textbf {\bibinfo {volume} {104}},\ \bibinfo {pages} {666} (\bibinfo
  {year} {1956})}\BibitemShut {NoStop}%
\bibitem [{\citenamefont {Semenenko}\ and\ \citenamefont
  {Esquinazi}(2018)}]{sem18}%
  \BibitemOpen
  \bibfield  {author} {\bibinfo {author} {\bibfnamefont {B.}~\bibnamefont
  {Semenenko}}\ and\ \bibinfo {author} {\bibfnamefont {P.}~\bibnamefont
  {Esquinazi}},\ }\href {\doibase 10.3390/magnetochemistry4040052} {\bibfield
  {journal} {\bibinfo  {journal} {Magnetochemistry}\ }\textbf {\bibinfo
  {volume} {4}},\ \bibinfo {pages} {52} (\bibinfo {year} {2018})}\BibitemShut
  {NoStop}%
\bibitem [{\citenamefont {McClure}(1957)}]{mcc1}%
  \BibitemOpen
  \bibfield  {author} {\bibinfo {author} {\bibfnamefont {J.~W.}\ \bibnamefont
  {McClure}},\ }\href@noop {} {\bibfield  {journal} {\bibinfo  {journal}
  {Physical Review}\ }\textbf {\bibinfo {volume} {108}},\ \bibinfo {pages}
  {612} (\bibinfo {year} {1957})}\BibitemShut {NoStop}%
\bibitem [{\citenamefont {Slonczewski}\ and\ \citenamefont
  {Weiss}(1958)}]{sw58}%
  \BibitemOpen
  \bibfield  {author} {\bibinfo {author} {\bibfnamefont {J.~C.}\ \bibnamefont
  {Slonczewski}}\ and\ \bibinfo {author} {\bibfnamefont {P.~R.}\ \bibnamefont
  {Weiss}},\ }\href@noop {} {\bibfield  {journal} {\bibinfo  {journal}
  {Physical Review}\ }\textbf {\bibinfo {volume} {109}},\ \bibinfo {pages}
  {272} (\bibinfo {year} {1958})}\BibitemShut {NoStop}%
\bibitem [{\citenamefont {McClure}(1958)}]{mcc2}%
  \BibitemOpen
  \bibfield  {author} {\bibinfo {author} {\bibfnamefont {J.~W.}\ \bibnamefont
  {McClure}},\ }\href@noop {} {\bibfield  {journal} {\bibinfo  {journal}
  {Physical Review}\ }\textbf {\bibinfo {volume} {112}},\ \bibinfo {pages}
  {715} (\bibinfo {year} {1958})}\BibitemShut {NoStop}%
\bibitem [{\citenamefont {Cao}\ \emph {et~al.}(2018)\citenamefont {Cao},
  \citenamefont {Fatemi}, \citenamefont {Fang}, \citenamefont {Watanabe},
  \citenamefont {Taniguchi}, \citenamefont {Kaxiras},\ and\ \citenamefont
  {Jarillo-Herrero}}]{cao18}%
  \BibitemOpen
  \bibfield  {author} {\bibinfo {author} {\bibfnamefont {Y.}~\bibnamefont
  {Cao}}, \bibinfo {author} {\bibfnamefont {V.}~\bibnamefont {Fatemi}},
  \bibinfo {author} {\bibfnamefont {S.}~\bibnamefont {Fang}}, \bibinfo {author}
  {\bibfnamefont {K.}~\bibnamefont {Watanabe}}, \bibinfo {author}
  {\bibfnamefont {T.}~\bibnamefont {Taniguchi}}, \bibinfo {author}
  {\bibfnamefont {E.}~\bibnamefont {Kaxiras}}, \ and\ \bibinfo {author}
  {\bibfnamefont {P.}~\bibnamefont {Jarillo-Herrero}},\ }\href {\doibase
  10.1038/nature26160} {\bibfield  {journal} {\bibinfo  {journal} {Nature}\
  }\textbf {\bibinfo {volume} {556}},\ \bibinfo {pages} {43} (\bibinfo {year}
  {2018})}\BibitemShut {NoStop}%
\bibitem [{\citenamefont {Marchenko}\ \emph {et~al.}(2018)\citenamefont
  {Marchenko}, \citenamefont {Evtushinsky}, \citenamefont {Golias},
  \citenamefont {Varykhalov}, \citenamefont {Seyller},\ and\ \citenamefont
  {Rader}}]{mar18}%
  \BibitemOpen
  \bibfield  {author} {\bibinfo {author} {\bibfnamefont {D.}~\bibnamefont
  {Marchenko}}, \bibinfo {author} {\bibfnamefont {D.~V.}\ \bibnamefont
  {Evtushinsky}}, \bibinfo {author} {\bibfnamefont {E.}~\bibnamefont {Golias}},
  \bibinfo {author} {\bibfnamefont {A.}~\bibnamefont {Varykhalov}}, \bibinfo
  {author} {\bibfnamefont {T.}~\bibnamefont {Seyller}}, \ and\ \bibinfo
  {author} {\bibfnamefont {O.}~\bibnamefont {Rader}},\ }\href {\doibase
  10.1126/sciadv.aau0059} {\bibfield  {journal} {\bibinfo  {journal} {Science
  Advances}\ }\textbf {\bibinfo {volume} {4}} (\bibinfo {year} {2018}),\
  10.1126/sciadv.aau0059}\BibitemShut {NoStop}%
\bibitem [{\citenamefont {Ballestar}\ \emph {et~al.}(2014)\citenamefont
  {Ballestar}, \citenamefont {Heikkil\"a},\ and\ \citenamefont
  {Esquinazi}}]{bal14I}%
  \BibitemOpen
  \bibfield  {author} {\bibinfo {author} {\bibfnamefont {A.}~\bibnamefont
  {Ballestar}}, \bibinfo {author} {\bibfnamefont {T.~T.}\ \bibnamefont
  {Heikkil\"a}}, \ and\ \bibinfo {author} {\bibfnamefont {P.}~\bibnamefont
  {Esquinazi}},\ }\href@noop {} {\bibfield  {journal} {\bibinfo  {journal}
  {Superc. Sci. Technol.}\ }\textbf {\bibinfo {volume} {27}},\ \bibinfo {pages}
  {115014} (\bibinfo {year} {2014})}\BibitemShut {NoStop}%
\bibitem [{\citenamefont {Esquinazi}\ \emph {et~al.}(2018)\citenamefont
  {Esquinazi}, \citenamefont {Precker}, \citenamefont {Stiller}, \citenamefont
  {Cordeiro}, \citenamefont {Barzola-Quiquia}, \citenamefont {Setzer},\ and\
  \citenamefont {B{\"o}hlmann}}]{esq18}%
  \BibitemOpen
  \bibfield  {author} {\bibinfo {author} {\bibfnamefont {P.~D.}\ \bibnamefont
  {Esquinazi}}, \bibinfo {author} {\bibfnamefont {C.~E.}\ \bibnamefont
  {Precker}}, \bibinfo {author} {\bibfnamefont {M.}~\bibnamefont {Stiller}},
  \bibinfo {author} {\bibfnamefont {T.~R.~S.}\ \bibnamefont {Cordeiro}},
  \bibinfo {author} {\bibfnamefont {J.}~\bibnamefont {Barzola-Quiquia}},
  \bibinfo {author} {\bibfnamefont {A.}~\bibnamefont {Setzer}}, \ and\ \bibinfo
  {author} {\bibfnamefont {W.}~\bibnamefont {B{\"o}hlmann}},\ }\href {\doibase
  10.1007/s40509-017-0131-0} {\bibfield  {journal} {\bibinfo  {journal}
  {Quantum Studies: Mathematics and Foundations}\ }\textbf {\bibinfo {volume}
  {5}},\ \bibinfo {pages} {41} (\bibinfo {year} {2018})}\BibitemShut {NoStop}%
\bibitem [{\citenamefont {Zoraghi}\ \emph {et~al.}(2017)\citenamefont
  {Zoraghi}, \citenamefont {Barzola-Quiquia}, \citenamefont {Stiller},
  \citenamefont {Setzer}, \citenamefont {Esquinazi}, \citenamefont {Kloess},
  \citenamefont {Muenster}, \citenamefont {L{\"u}hmann},\ and\ \citenamefont
  {Estrela-Lopis}}]{zor17}%
  \BibitemOpen
  \bibfield  {author} {\bibinfo {author} {\bibfnamefont {M.}~\bibnamefont
  {Zoraghi}}, \bibinfo {author} {\bibfnamefont {J.}~\bibnamefont
  {Barzola-Quiquia}}, \bibinfo {author} {\bibfnamefont {M.}~\bibnamefont
  {Stiller}}, \bibinfo {author} {\bibfnamefont {A.}~\bibnamefont {Setzer}},
  \bibinfo {author} {\bibfnamefont {P.}~\bibnamefont {Esquinazi}}, \bibinfo
  {author} {\bibfnamefont {G.}~\bibnamefont {Kloess}}, \bibinfo {author}
  {\bibfnamefont {T.}~\bibnamefont {Muenster}}, \bibinfo {author}
  {\bibfnamefont {T.}~\bibnamefont {L{\"u}hmann}}, \ and\ \bibinfo {author}
  {\bibfnamefont {I.}~\bibnamefont {Estrela-Lopis}},\ }\href {\doibase
  10.1103/PhysRevB.95.045308} {\bibfield  {journal} {\bibinfo  {journal} {Phys.
  Rev. B}\ }\textbf {\bibinfo {volume} {95}},\ \bibinfo {pages} {045308}
  (\bibinfo {year} {2017})}\BibitemShut {NoStop}%
\bibitem [{\citenamefont {Barzola-Quiquia}\ \emph {et~al.}(2019)\citenamefont
  {Barzola-Quiquia}, \citenamefont {Esquinazi}, \citenamefont {Precker},
  \citenamefont {Stiller}, \citenamefont {Zoraghi}, \citenamefont {F\"orster},
  \citenamefont {Herrmannsd\"orfer},\ and\ \citenamefont {Coniglio}}]{bar19}%
  \BibitemOpen
  \bibfield  {author} {\bibinfo {author} {\bibfnamefont {J.}~\bibnamefont
  {Barzola-Quiquia}}, \bibinfo {author} {\bibfnamefont {P.~D.}\ \bibnamefont
  {Esquinazi}}, \bibinfo {author} {\bibfnamefont {C.~E.}\ \bibnamefont
  {Precker}}, \bibinfo {author} {\bibfnamefont {M.}~\bibnamefont {Stiller}},
  \bibinfo {author} {\bibfnamefont {M.}~\bibnamefont {Zoraghi}}, \bibinfo
  {author} {\bibfnamefont {T.}~\bibnamefont {F\"orster}}, \bibinfo {author}
  {\bibfnamefont {T.}~\bibnamefont {Herrmannsd\"orfer}}, \ and\ \bibinfo
  {author} {\bibfnamefont {W.~A.}\ \bibnamefont {Coniglio}},\ }\href {\doibase
  10.1103/PhysRevMaterials.3.054603} {\bibfield  {journal} {\bibinfo  {journal}
  {Phys. Rev. Materials}\ }\textbf {\bibinfo {volume} {3}},\ \bibinfo {pages}
  {054603} (\bibinfo {year} {2019})}\BibitemShut {NoStop}%
\bibitem [{\citenamefont {Kinchin}(1953)}]{kin53}%
  \BibitemOpen
  \bibfield  {author} {\bibinfo {author} {\bibfnamefont {G.~H.}\ \bibnamefont
  {Kinchin}},\ }\href@noop {} {\bibfield  {journal} {\bibinfo  {journal} {Proc.
  R. Soc. Lond. A}\ }\textbf {\bibinfo {volume} {217}},\ \bibinfo {pages}
  {1128} (\bibinfo {year} {1953})}\BibitemShut {NoStop}%
\bibitem [{\citenamefont {Mrozowski}\ and\ \citenamefont
  {Chaberski}(1956)}]{mro56}%
  \BibitemOpen
  \bibfield  {author} {\bibinfo {author} {\bibfnamefont {S.}~\bibnamefont
  {Mrozowski}}\ and\ \bibinfo {author} {\bibfnamefont {A.}~\bibnamefont
  {Chaberski}},\ }\href@noop {} {\bibfield  {journal} {\bibinfo  {journal}
  {Phys. Rev.}\ }\textbf {\bibinfo {volume} {104}},\ \bibinfo {pages} {74}
  (\bibinfo {year} {1956})}\BibitemShut {NoStop}%
\bibitem [{\citenamefont {Soule}(1958)}]{sou58}%
  \BibitemOpen
  \bibfield  {author} {\bibinfo {author} {\bibfnamefont {D.~E.}\ \bibnamefont
  {Soule}},\ }\href@noop {} {\bibfield  {journal} {\bibinfo  {journal} {Phys.
  Rev.}\ }\textbf {\bibinfo {volume} {112}},\ \bibinfo {pages} {698} (\bibinfo
  {year} {1958})}\BibitemShut {NoStop}%
\bibitem [{\citenamefont {Cooper}\ \emph {et~al.}(1970)\citenamefont {Cooper},
  \citenamefont {Woore},\ and\ \citenamefont {Young}}]{coo70}%
  \BibitemOpen
  \bibfield  {author} {\bibinfo {author} {\bibfnamefont {J.~D.}\ \bibnamefont
  {Cooper}}, \bibinfo {author} {\bibfnamefont {J.}~\bibnamefont {Woore}}, \
  and\ \bibinfo {author} {\bibfnamefont {D.~A.}\ \bibnamefont {Young}},\
  }\href@noop {} {\bibfield  {journal} {\bibinfo  {journal} {Nature}\ }\textbf
  {\bibinfo {volume} {721--722}},\ \bibinfo {pages} {225} (\bibinfo {year}
  {1970})}\BibitemShut {NoStop}%
\bibitem [{\citenamefont {Brandt}\ \emph {et~al.}(1974)\citenamefont {Brandt},
  \citenamefont {Kapustin}, \citenamefont {Karavaev}, \citenamefont
  {Kotosonov},\ and\ \citenamefont {Svistova}}]{bra74}%
  \BibitemOpen
  \bibfield  {author} {\bibinfo {author} {\bibfnamefont {N.~B.}\ \bibnamefont
  {Brandt}}, \bibinfo {author} {\bibfnamefont {G.~A.}\ \bibnamefont
  {Kapustin}}, \bibinfo {author} {\bibfnamefont {V.~G.}\ \bibnamefont
  {Karavaev}}, \bibinfo {author} {\bibfnamefont {A.~S.}\ \bibnamefont
  {Kotosonov}}, \ and\ \bibinfo {author} {\bibfnamefont {E.~A.}\ \bibnamefont
  {Svistova}},\ }\href@noop {} {\bibfield  {journal} {\bibinfo  {journal} {Sov.
  Phys.-JETP}\ }\textbf {\bibinfo {volume} {40}},\ \bibinfo {pages} {564}
  (\bibinfo {year} {1974})}\BibitemShut {NoStop}%
\bibitem [{\citenamefont {Oshima}\ \emph {et~al.}(1982)\citenamefont {Oshima},
  \citenamefont {Kawamura}, \citenamefont {Tsuzuku},\ and\ \citenamefont
  {Sugihara}}]{osh82}%
  \BibitemOpen
  \bibfield  {author} {\bibinfo {author} {\bibfnamefont {H.}~\bibnamefont
  {Oshima}}, \bibinfo {author} {\bibfnamefont {K.}~\bibnamefont {Kawamura}},
  \bibinfo {author} {\bibfnamefont {T.}~\bibnamefont {Tsuzuku}}, \ and\
  \bibinfo {author} {\bibfnamefont {K.}~\bibnamefont {Sugihara}},\ }\href@noop
  {} {\bibfield  {journal} {\bibinfo  {journal} {J Phys. Soc. Japan}\ }\textbf
  {\bibinfo {volume} {51}},\ \bibinfo {pages} {1476} (\bibinfo {year}
  {1982})}\BibitemShut {NoStop}%
\bibitem [{\citenamefont {Bunch}\ \emph {et~al.}(2005)\citenamefont {Bunch},
  \citenamefont {Yaish}, \citenamefont {Brink}, \citenamefont {Bolotin},\ and\
  \citenamefont {McEuen}}]{bun05}%
  \BibitemOpen
  \bibfield  {author} {\bibinfo {author} {\bibfnamefont {J.~S.}\ \bibnamefont
  {Bunch}}, \bibinfo {author} {\bibfnamefont {Y.}~\bibnamefont {Yaish}},
  \bibinfo {author} {\bibfnamefont {M.}~\bibnamefont {Brink}}, \bibinfo
  {author} {\bibfnamefont {K.}~\bibnamefont {Bolotin}}, \ and\ \bibinfo
  {author} {\bibfnamefont {P.~L.}\ \bibnamefont {McEuen}},\ }\href@noop {}
  {\bibfield  {journal} {\bibinfo  {journal} {Nano Letters}\ }\textbf {\bibinfo
  {volume} {5}},\ \bibinfo {pages} {287} (\bibinfo {year} {2005})}\BibitemShut
  {NoStop}%
\bibitem [{\citenamefont {Vansweevelt}\ \emph {et~al.}(2011)\citenamefont
  {Vansweevelt}, \citenamefont {Mortet}, \citenamefont {D'Haen}, \citenamefont
  {Ruttens}, \citenamefont {Haesendonck}, \citenamefont {Partoens},
  \citenamefont {Peeters},\ and\ \citenamefont {Wagner}}]{van11}%
  \BibitemOpen
  \bibfield  {author} {\bibinfo {author} {\bibfnamefont {R.}~\bibnamefont
  {Vansweevelt}}, \bibinfo {author} {\bibfnamefont {V.}~\bibnamefont {Mortet}},
  \bibinfo {author} {\bibfnamefont {J.}~\bibnamefont {D'Haen}}, \bibinfo
  {author} {\bibfnamefont {B.}~\bibnamefont {Ruttens}}, \bibinfo {author}
  {\bibfnamefont {C.~V.}\ \bibnamefont {Haesendonck}}, \bibinfo {author}
  {\bibfnamefont {B.}~\bibnamefont {Partoens}}, \bibinfo {author}
  {\bibfnamefont {F.~M.}\ \bibnamefont {Peeters}}, \ and\ \bibinfo {author}
  {\bibfnamefont {P.}~\bibnamefont {Wagner}},\ }\href@noop {} {\bibfield
  {journal} {\bibinfo  {journal} {Phys. Status Solidi A}\ }\textbf {\bibinfo
  {volume} {208}},\ \bibinfo {pages} {1252} (\bibinfo {year}
  {2011})}\BibitemShut {NoStop}%
\bibitem [{\citenamefont {Kopelevich}\ \emph {et~al.}(2003)\citenamefont
  {Kopelevich}, \citenamefont {Torres}, \citenamefont {da~Silva}, \citenamefont
  {Mrowka}, \citenamefont {Kempa},\ and\ \citenamefont
  {Esquinazi}}]{yakovprl03}%
  \BibitemOpen
  \bibfield  {author} {\bibinfo {author} {\bibfnamefont {Y.}~\bibnamefont
  {Kopelevich}}, \bibinfo {author} {\bibfnamefont {J.~H.~S.}\ \bibnamefont
  {Torres}}, \bibinfo {author} {\bibfnamefont {R.~R.}\ \bibnamefont
  {da~Silva}}, \bibinfo {author} {\bibfnamefont {F.}~\bibnamefont {Mrowka}},
  \bibinfo {author} {\bibfnamefont {H.}~\bibnamefont {Kempa}}, \ and\ \bibinfo
  {author} {\bibfnamefont {P.}~\bibnamefont {Esquinazi}},\ }\href@noop {}
  {\bibfield  {journal} {\bibinfo  {journal} {Phys. Rev. Lett.}\ }\textbf
  {\bibinfo {volume} {90}},\ \bibinfo {pages} {156402} (\bibinfo {year}
  {2003})}\BibitemShut {NoStop}%
\bibitem [{\citenamefont {Kempa}\ \emph {et~al.}(2006)\citenamefont {Kempa},
  \citenamefont {Esquinazi},\ and\ \citenamefont {Kopelevich}}]{kempa06}%
  \BibitemOpen
  \bibfield  {author} {\bibinfo {author} {\bibfnamefont {H.}~\bibnamefont
  {Kempa}}, \bibinfo {author} {\bibfnamefont {P.}~\bibnamefont {Esquinazi}}, \
  and\ \bibinfo {author} {\bibfnamefont {Y.}~\bibnamefont {Kopelevich}},\
  }\href@noop {} {\bibfield  {journal} {\bibinfo  {journal} {Solid State
  Communications}\ }\textbf {\bibinfo {volume} {138}},\ \bibinfo {pages} {118}
  (\bibinfo {year} {2006})}\BibitemShut {NoStop}%
\bibitem [{\citenamefont {Kopelevich}\ \emph {et~al.}(2006)\citenamefont
  {Kopelevich}, \citenamefont {Pantoja}, \citenamefont {da~Silva},
  \citenamefont {Mrowka},\ and\ \citenamefont {Esquinazi}}]{kopepl06}%
  \BibitemOpen
  \bibfield  {author} {\bibinfo {author} {\bibfnamefont {Y.}~\bibnamefont
  {Kopelevich}}, \bibinfo {author} {\bibfnamefont {J.~M.}\ \bibnamefont
  {Pantoja}}, \bibinfo {author} {\bibfnamefont {R.}~\bibnamefont {da~Silva}},
  \bibinfo {author} {\bibfnamefont {F.}~\bibnamefont {Mrowka}}, \ and\ \bibinfo
  {author} {\bibfnamefont {P.}~\bibnamefont {Esquinazi}},\ }\href@noop {}
  {\bibfield  {journal} {\bibinfo  {journal} {Phys. Lett. A}\ }\textbf
  {\bibinfo {volume} {355}},\ \bibinfo {pages} {233} (\bibinfo {year}
  {2006})}\BibitemShut {NoStop}%
\bibitem [{\citenamefont {Schneider}\ \emph {et~al.}(2009)\citenamefont
  {Schneider}, \citenamefont {Orlita}, \citenamefont {Potemski},\ and\
  \citenamefont {Maude}}]{sch09}%
  \BibitemOpen
  \bibfield  {author} {\bibinfo {author} {\bibfnamefont {J.~M.}\ \bibnamefont
  {Schneider}}, \bibinfo {author} {\bibfnamefont {M.}~\bibnamefont {Orlita}},
  \bibinfo {author} {\bibfnamefont {M.}~\bibnamefont {Potemski}}, \ and\
  \bibinfo {author} {\bibfnamefont {D.~K.}\ \bibnamefont {Maude}},\ }\href@noop
  {} {\bibfield  {journal} {\bibinfo  {journal} {Phys. Rev. Lett.}\ }\textbf
  {\bibinfo {volume} {102}},\ \bibinfo {pages} {166403} (\bibinfo {year}
  {2009})},\ \bibinfo {note} {\rm see also the comment by I. A. Luk'yanchuk and
  Y. Kopelevich, idem {\bf 104}, 119701 (2010).}\BibitemShut {Stop}%
\bibitem [{\citenamefont {Esquinazi}\ \emph {et~al.}(2014)\citenamefont
  {Esquinazi}, \citenamefont {Kr\"uger}, \citenamefont {Barzola-Quiquia},
  \citenamefont {Sch\"onemann}, \citenamefont {Hermannsd\"orfer},\ and\
  \citenamefont {Garc\'ia}}]{esq14}%
  \BibitemOpen
  \bibfield  {author} {\bibinfo {author} {\bibfnamefont {P.}~\bibnamefont
  {Esquinazi}}, \bibinfo {author} {\bibfnamefont {J.}~\bibnamefont {Kr\"uger}},
  \bibinfo {author} {\bibfnamefont {J.}~\bibnamefont {Barzola-Quiquia}},
  \bibinfo {author} {\bibfnamefont {R.}~\bibnamefont {Sch\"onemann}}, \bibinfo
  {author} {\bibfnamefont {T.}~\bibnamefont {Hermannsd\"orfer}}, \ and\
  \bibinfo {author} {\bibfnamefont {N.}~\bibnamefont {Garc\'ia}},\ }\href@noop
  {} {\bibfield  {journal} {\bibinfo  {journal} {AIP Advances}\ }\textbf
  {\bibinfo {volume} {4}},\ \bibinfo {pages} {117121} (\bibinfo {year}
  {2014})}\BibitemShut {NoStop}%
\bibitem [{\citenamefont {Breusing}\ \emph {et~al.}(2009)\citenamefont
  {Breusing}, \citenamefont {Ropers},\ and\ \citenamefont {Elsaesser}}]{bre09}%
  \BibitemOpen
  \bibfield  {author} {\bibinfo {author} {\bibfnamefont {M.}~\bibnamefont
  {Breusing}}, \bibinfo {author} {\bibfnamefont {C.}~\bibnamefont {Ropers}}, \
  and\ \bibinfo {author} {\bibfnamefont {T.}~\bibnamefont {Elsaesser}},\ }\href
  {\doibase 10.1103/PhysRevLett.102.086809} {\bibfield  {journal} {\bibinfo
  {journal} {Phys. Rev. Lett.}\ }\textbf {\bibinfo {volume} {102}},\ \bibinfo
  {pages} {086809} (\bibinfo {year} {2009})}\BibitemShut {NoStop}%
\bibitem [{\citenamefont {Sugawara}\ \emph {et~al.}(2007)\citenamefont
  {Sugawara}, \citenamefont {Sato}, \citenamefont {Souma}, \citenamefont
  {Takahashi},\ and\ \citenamefont {Suematsu}}]{sug07}%
  \BibitemOpen
  \bibfield  {author} {\bibinfo {author} {\bibfnamefont {K.}~\bibnamefont
  {Sugawara}}, \bibinfo {author} {\bibfnamefont {T.}~\bibnamefont {Sato}},
  \bibinfo {author} {\bibfnamefont {S.}~\bibnamefont {Souma}}, \bibinfo
  {author} {\bibfnamefont {T.}~\bibnamefont {Takahashi}}, \ and\ \bibinfo
  {author} {\bibfnamefont {H.}~\bibnamefont {Suematsu}},\ }\href {\doibase
  10.1103/PhysRevLett.98.036801} {\bibfield  {journal} {\bibinfo  {journal}
  {Phys. Rev. Lett.}\ }\textbf {\bibinfo {volume} {98}},\ \bibinfo {pages}
  {036801} (\bibinfo {year} {2007})}\BibitemShut {NoStop}%
\bibitem [{\citenamefont {Gr\"uneis}\ \emph {et~al.}(2008)\citenamefont
  {Gr\"uneis}, \citenamefont {Attaccalite}, \citenamefont {Pichler},
  \citenamefont {Zabolotnyy}, \citenamefont {Shiozawa}, \citenamefont
  {Molodtsov}, \citenamefont {Inosov}, \citenamefont {Koitzsch}, \citenamefont
  {Knupfer}, \citenamefont {Schiessling}, \citenamefont {Follath},
  \citenamefont {Weber}, \citenamefont {Rudolf}, \citenamefont {Wirtz},\ and\
  \citenamefont {Rubio}}]{gru08}%
  \BibitemOpen
  \bibfield  {author} {\bibinfo {author} {\bibfnamefont {A.}~\bibnamefont
  {Gr\"uneis}}, \bibinfo {author} {\bibfnamefont {C.}~\bibnamefont
  {Attaccalite}}, \bibinfo {author} {\bibfnamefont {T.}~\bibnamefont
  {Pichler}}, \bibinfo {author} {\bibfnamefont {V.}~\bibnamefont {Zabolotnyy}},
  \bibinfo {author} {\bibfnamefont {H.}~\bibnamefont {Shiozawa}}, \bibinfo
  {author} {\bibfnamefont {S.~L.}\ \bibnamefont {Molodtsov}}, \bibinfo {author}
  {\bibfnamefont {D.}~\bibnamefont {Inosov}}, \bibinfo {author} {\bibfnamefont
  {A.}~\bibnamefont {Koitzsch}}, \bibinfo {author} {\bibfnamefont
  {M.}~\bibnamefont {Knupfer}}, \bibinfo {author} {\bibfnamefont
  {J.}~\bibnamefont {Schiessling}}, \bibinfo {author} {\bibfnamefont
  {R.}~\bibnamefont {Follath}}, \bibinfo {author} {\bibfnamefont
  {R.}~\bibnamefont {Weber}}, \bibinfo {author} {\bibfnamefont
  {P.}~\bibnamefont {Rudolf}}, \bibinfo {author} {\bibfnamefont
  {R.}~\bibnamefont {Wirtz}}, \ and\ \bibinfo {author} {\bibfnamefont
  {A.}~\bibnamefont {Rubio}},\ }\href {\doibase 10.1103/PhysRevLett.100.037601}
  {\bibfield  {journal} {\bibinfo  {journal} {Phys. Rev. Lett.}\ }\textbf
  {\bibinfo {volume} {100}},\ \bibinfo {pages} {037601} (\bibinfo {year}
  {2008})}\BibitemShut {NoStop}%
\bibitem [{\citenamefont {Liu}\ \emph {et~al.}(2010)\citenamefont {Liu},
  \citenamefont {Zhang}, \citenamefont {Brinkley}, \citenamefont {Bian},
  \citenamefont {Miller},\ and\ \citenamefont {Chiang}}]{liu10}%
  \BibitemOpen
  \bibfield  {author} {\bibinfo {author} {\bibfnamefont {Y.}~\bibnamefont
  {Liu}}, \bibinfo {author} {\bibfnamefont {L.}~\bibnamefont {Zhang}}, \bibinfo
  {author} {\bibfnamefont {M.~K.}\ \bibnamefont {Brinkley}}, \bibinfo {author}
  {\bibfnamefont {G.}~\bibnamefont {Bian}}, \bibinfo {author} {\bibfnamefont
  {T.}~\bibnamefont {Miller}}, \ and\ \bibinfo {author} {\bibfnamefont {T.-C.}\
  \bibnamefont {Chiang}},\ }\href {\doibase 10.1103/PhysRevLett.105.136804}
  {\bibfield  {journal} {\bibinfo  {journal} {Phys. Rev. Lett.}\ }\textbf
  {\bibinfo {volume} {105}},\ \bibinfo {pages} {136804} (\bibinfo {year}
  {2010})}\BibitemShut {NoStop}%
\bibitem [{\citenamefont {Yin}\ \emph {et~al.}(2020)\citenamefont {Yin},
  \citenamefont {Zheng}, \citenamefont {Ma}, \citenamefont {Liao},
  \citenamefont {Ma},\ and\ \citenamefont {Wang}}]{yin20}%
  \BibitemOpen
  \bibfield  {author} {\bibinfo {author} {\bibfnamefont {R.}~\bibnamefont
  {Yin}}, \bibinfo {author} {\bibfnamefont {Y.}~\bibnamefont {Zheng}}, \bibinfo
  {author} {\bibfnamefont {X.}~\bibnamefont {Ma}}, \bibinfo {author}
  {\bibfnamefont {Q.}~\bibnamefont {Liao}}, \bibinfo {author} {\bibfnamefont
  {C.}~\bibnamefont {Ma}}, \ and\ \bibinfo {author} {\bibfnamefont
  {B.}~\bibnamefont {Wang}},\ }\href {\doibase 10.1103/PhysRevB.102.115410}
  {\bibfield  {journal} {\bibinfo  {journal} {Phys. Rev. B}\ }\textbf {\bibinfo
  {volume} {102}},\ \bibinfo {pages} {115410} (\bibinfo {year}
  {2020})}\BibitemShut {NoStop}%
\bibitem [{\citenamefont {Gonz{\'a}lez}\ \emph {et~al.}(2007)\citenamefont
  {Gonz{\'a}lez}, \citenamefont {Mu{\~n}oz}, \citenamefont {Garc\'ia},
  \citenamefont {Barzola-Quiquia}, \citenamefont {Spoddig}, \citenamefont
  {Schindler},\ and\ \citenamefont {Esquinazi}}]{gon07}%
  \BibitemOpen
  \bibfield  {author} {\bibinfo {author} {\bibfnamefont {J.~C.}\ \bibnamefont
  {Gonz{\'a}lez}}, \bibinfo {author} {\bibfnamefont {M.}~\bibnamefont
  {Mu{\~n}oz}}, \bibinfo {author} {\bibfnamefont {N.}~\bibnamefont {Garc\'ia}},
  \bibinfo {author} {\bibfnamefont {J.}~\bibnamefont {Barzola-Quiquia}},
  \bibinfo {author} {\bibfnamefont {D.}~\bibnamefont {Spoddig}}, \bibinfo
  {author} {\bibfnamefont {K.}~\bibnamefont {Schindler}}, \ and\ \bibinfo
  {author} {\bibfnamefont {P.}~\bibnamefont {Esquinazi}},\ }\href@noop {}
  {\bibfield  {journal} {\bibinfo  {journal} {Phys. Rev. Lett.}\ }\textbf
  {\bibinfo {volume} {99}},\ \bibinfo {pages} {216601} (\bibinfo {year}
  {2007})}\BibitemShut {NoStop}%
\bibitem [{\citenamefont {Henck}\ \emph {et~al.}(2018)\citenamefont {Henck},
  \citenamefont {Avila}, \citenamefont {Ben~Aziza}, \citenamefont {Pierucci},
  \citenamefont {Baima}, \citenamefont {Pamuk}, \citenamefont {Chaste},
  \citenamefont {Utt}, \citenamefont {Bartos}, \citenamefont {Nogajewski},
  \citenamefont {Piot}, \citenamefont {Orlita}, \citenamefont {Potemski},
  \citenamefont {Calandra}, \citenamefont {Asensio}, \citenamefont {Mauri},
  \citenamefont {Faugeras},\ and\ \citenamefont {Ouerghi}}]{hen18}%
  \BibitemOpen
  \bibfield  {author} {\bibinfo {author} {\bibfnamefont {H.}~\bibnamefont
  {Henck}}, \bibinfo {author} {\bibfnamefont {J.}~\bibnamefont {Avila}},
  \bibinfo {author} {\bibfnamefont {Z.}~\bibnamefont {Ben~Aziza}}, \bibinfo
  {author} {\bibfnamefont {D.}~\bibnamefont {Pierucci}}, \bibinfo {author}
  {\bibfnamefont {J.}~\bibnamefont {Baima}}, \bibinfo {author} {\bibfnamefont
  {B.}~\bibnamefont {Pamuk}}, \bibinfo {author} {\bibfnamefont
  {J.}~\bibnamefont {Chaste}}, \bibinfo {author} {\bibfnamefont
  {D.}~\bibnamefont {Utt}}, \bibinfo {author} {\bibfnamefont {M.}~\bibnamefont
  {Bartos}}, \bibinfo {author} {\bibfnamefont {K.}~\bibnamefont {Nogajewski}},
  \bibinfo {author} {\bibfnamefont {B.~A.}\ \bibnamefont {Piot}}, \bibinfo
  {author} {\bibfnamefont {M.}~\bibnamefont {Orlita}}, \bibinfo {author}
  {\bibfnamefont {M.}~\bibnamefont {Potemski}}, \bibinfo {author}
  {\bibfnamefont {M.}~\bibnamefont {Calandra}}, \bibinfo {author}
  {\bibfnamefont {M.~C.}\ \bibnamefont {Asensio}}, \bibinfo {author}
  {\bibfnamefont {F.}~\bibnamefont {Mauri}}, \bibinfo {author} {\bibfnamefont
  {C.}~\bibnamefont {Faugeras}}, \ and\ \bibinfo {author} {\bibfnamefont
  {A.}~\bibnamefont {Ouerghi}},\ }\href {\doibase 10.1103/PhysRevB.97.245421}
  {\bibfield  {journal} {\bibinfo  {journal} {Phys. Rev. B}\ }\textbf {\bibinfo
  {volume} {97}},\ \bibinfo {pages} {245421} (\bibinfo {year}
  {2018})}\BibitemShut {NoStop}%
\bibitem [{\citenamefont {Li}\ \emph {et~al.}(2011)\citenamefont {Li},
  \citenamefont {Minne}, \citenamefont {Pittenger},\ and\ \citenamefont
  {Mednick}}]{PF}%
  \BibitemOpen
  \bibfield  {author} {\bibinfo {author} {\bibfnamefont {C.}~\bibnamefont
  {Li}}, \bibinfo {author} {\bibfnamefont {S.}~\bibnamefont {Minne}}, \bibinfo
  {author} {\bibfnamefont {B.}~\bibnamefont {Pittenger}}, \ and\ \bibinfo
  {author} {\bibfnamefont {A.}~\bibnamefont {Mednick}},\ }\href@noop {}
  {\enquote {\bibinfo {title} {{Simultaneous Electrical and Mechanical Property
  Mapping at the Nanoscale with PeakForce TUNA}},}\ } (\bibinfo {year}
  {2011}),\ \bibinfo {note} {bruker Application Note \#132 (2011)}\BibitemShut
  {NoStop}%
\bibitem [{\citenamefont {Pierucci}\ \emph {et~al.}(2015)\citenamefont
  {Pierucci}, \citenamefont {Sediri}, \citenamefont {Hajlaoui}, \citenamefont
  {Girard}, \citenamefont {Brumme}, \citenamefont {Calandra}, \citenamefont
  {Velez-Fort}, \citenamefont {Patriarche}, \citenamefont {Silly},
  \citenamefont {Ferro}, \citenamefont {Souliere}, \citenamefont {Marangolo},
  \citenamefont {Sirotti}, \citenamefont {Mauri},\ and\ \citenamefont
  {Ouerghi}}]{pie15}%
  \BibitemOpen
  \bibfield  {author} {\bibinfo {author} {\bibfnamefont {D.}~\bibnamefont
  {Pierucci}}, \bibinfo {author} {\bibfnamefont {H.}~\bibnamefont {Sediri}},
  \bibinfo {author} {\bibfnamefont {M.}~\bibnamefont {Hajlaoui}}, \bibinfo
  {author} {\bibfnamefont {J.-C.}\ \bibnamefont {Girard}}, \bibinfo {author}
  {\bibfnamefont {T.}~\bibnamefont {Brumme}}, \bibinfo {author} {\bibfnamefont
  {M.}~\bibnamefont {Calandra}}, \bibinfo {author} {\bibfnamefont
  {E.}~\bibnamefont {Velez-Fort}}, \bibinfo {author} {\bibfnamefont
  {G.}~\bibnamefont {Patriarche}}, \bibinfo {author} {\bibfnamefont {M.~G.}\
  \bibnamefont {Silly}}, \bibinfo {author} {\bibfnamefont {G.}~\bibnamefont
  {Ferro}}, \bibinfo {author} {\bibfnamefont {V.}~\bibnamefont {Souliere}},
  \bibinfo {author} {\bibfnamefont {M.}~\bibnamefont {Marangolo}}, \bibinfo
  {author} {\bibfnamefont {F.}~\bibnamefont {Sirotti}}, \bibinfo {author}
  {\bibfnamefont {F.}~\bibnamefont {Mauri}}, \ and\ \bibinfo {author}
  {\bibfnamefont {A.}~\bibnamefont {Ouerghi}},\ }\href {\doibase
  10.1021/acsnano.5b01239} {\bibfield  {journal} {\bibinfo  {journal} {ACS
  Nano}\ }\textbf {\bibinfo {volume} {9}},\ \bibinfo {pages} {5432} (\bibinfo
  {year} {2015})}\BibitemShut {NoStop}%
\bibitem [{\citenamefont {Kopnin}\ \emph {et~al.}(2013)\citenamefont {Kopnin},
  \citenamefont {Ij\"as}, \citenamefont {Harju},\ and\ \citenamefont
  {Heikkil\"a}}]{kop13}%
  \BibitemOpen
  \bibfield  {author} {\bibinfo {author} {\bibfnamefont {N.~B.}\ \bibnamefont
  {Kopnin}}, \bibinfo {author} {\bibfnamefont {M.}~\bibnamefont {Ij\"as}},
  \bibinfo {author} {\bibfnamefont {A.}~\bibnamefont {Harju}}, \ and\ \bibinfo
  {author} {\bibfnamefont {T.~T.}\ \bibnamefont {Heikkil\"a}},\ }\href
  {\doibase 10.1103/PhysRevB.87.140503} {\bibfield  {journal} {\bibinfo
  {journal} {Phys. Rev. B}\ }\textbf {\bibinfo {volume} {87}},\ \bibinfo
  {pages} {140503} (\bibinfo {year} {2013})}\BibitemShut {NoStop}%
\bibitem [{\citenamefont {Volovik}(2018)}]{vol18}%
  \BibitemOpen
  \bibfield  {author} {\bibinfo {author} {\bibfnamefont {G.~E.}\ \bibnamefont
  {Volovik}},\ }\href {https://doi.org/10.1134/S0021364018080052} {\bibfield
  {journal} {\bibinfo  {journal} {JETP Letters}\ }\textbf {\bibinfo {volume}
  {107}},\ \bibinfo {pages} {516} (\bibinfo {year} {2018})}\BibitemShut
  {NoStop}%
\bibitem [{\citenamefont {Cong}\ \emph {et~al.}(2011)\citenamefont {Cong},
  \citenamefont {Yu}, \citenamefont {Sato}, \citenamefont {Shang},
  \citenamefont {Saito}, \citenamefont {Dresselhaus},\ and\ \citenamefont
  {Dresselhaus}}]{con11}%
  \BibitemOpen
  \bibfield  {author} {\bibinfo {author} {\bibfnamefont {C.}~\bibnamefont
  {Cong}}, \bibinfo {author} {\bibfnamefont {T.}~\bibnamefont {Yu}}, \bibinfo
  {author} {\bibfnamefont {K.}~\bibnamefont {Sato}}, \bibinfo {author}
  {\bibfnamefont {J.}~\bibnamefont {Shang}}, \bibinfo {author} {\bibfnamefont
  {R.}~\bibnamefont {Saito}}, \bibinfo {author} {\bibfnamefont {G.~F.}\
  \bibnamefont {Dresselhaus}}, \ and\ \bibinfo {author} {\bibfnamefont {M.~S.}\
  \bibnamefont {Dresselhaus}},\ }\href {\doibase 10.1021/nn203472f} {\bibfield
  {journal} {\bibinfo  {journal} {ACS Nano}\ }\textbf {\bibinfo {volume} {5}},\
  \bibinfo {pages} {8760} (\bibinfo {year} {2011})},\ \Eprint
  {http://arxiv.org/abs/https://doi.org/10.1021/nn203472f}
  {https://doi.org/10.1021/nn203472f} \BibitemShut {NoStop}%
\bibitem [{\citenamefont {Henni}\ \emph {et~al.}(2016)\citenamefont {Henni},
  \citenamefont {Collado}, \citenamefont {Nogajewski}, \citenamefont {Molas},
  \citenamefont {Usaj}, \citenamefont {Balseiro}, \citenamefont {Orlita},
  \citenamefont {Potemski},\ and\ \citenamefont {Faugeras}}]{hen16}%
  \BibitemOpen
  \bibfield  {author} {\bibinfo {author} {\bibfnamefont {Y.}~\bibnamefont
  {Henni}}, \bibinfo {author} {\bibfnamefont {H.~P.~O.}\ \bibnamefont
  {Collado}}, \bibinfo {author} {\bibfnamefont {K.}~\bibnamefont {Nogajewski}},
  \bibinfo {author} {\bibfnamefont {M.~R.}\ \bibnamefont {Molas}}, \bibinfo
  {author} {\bibfnamefont {G.}~\bibnamefont {Usaj}}, \bibinfo {author}
  {\bibfnamefont {C.~A.}\ \bibnamefont {Balseiro}}, \bibinfo {author}
  {\bibfnamefont {M.}~\bibnamefont {Orlita}}, \bibinfo {author} {\bibfnamefont
  {M.}~\bibnamefont {Potemski}}, \ and\ \bibinfo {author} {\bibfnamefont
  {C.}~\bibnamefont {Faugeras}},\ }\href@noop {} {\bibfield  {journal}
  {\bibinfo  {journal} {Nano Letters}\ }\textbf {\bibinfo {volume} {16}},\
  \bibinfo {pages} {3710} (\bibinfo {year} {2016})}\BibitemShut {NoStop}%
\bibitem [{\citenamefont {Torche}\ \emph {et~al.}(2017)\citenamefont {Torche},
  \citenamefont {Mauri}, \citenamefont {Charlier},\ and\ \citenamefont
  {Calandra}}]{tor17}%
  \BibitemOpen
  \bibfield  {author} {\bibinfo {author} {\bibfnamefont {A.}~\bibnamefont
  {Torche}}, \bibinfo {author} {\bibfnamefont {F.}~\bibnamefont {Mauri}},
  \bibinfo {author} {\bibfnamefont {J.-C.}\ \bibnamefont {Charlier}}, \ and\
  \bibinfo {author} {\bibfnamefont {M.}~\bibnamefont {Calandra}},\ }\href
  {\doibase 10.1103/PhysRevMaterials.1.041001} {\bibfield  {journal} {\bibinfo
  {journal} {Phys. Rev. Materials}\ }\textbf {\bibinfo {volume} {1}},\ \bibinfo
  {pages} {041001} (\bibinfo {year} {2017})}\BibitemShut {NoStop}%
\bibitem [{\citenamefont {Ramos}\ \emph {et~al.}(2021)\citenamefont {Ramos},
  \citenamefont {Pimenta},\ and\ \citenamefont {Champi}}]{ram20}%
  \BibitemOpen
  \bibfield  {author} {\bibinfo {author} {\bibfnamefont {S.~L.~L.}\
  \bibnamefont {Ramos}}, \bibinfo {author} {\bibfnamefont {M.~A.}\ \bibnamefont
  {Pimenta}}, \ and\ \bibinfo {author} {\bibfnamefont {A.}~\bibnamefont
  {Champi}},\ }\href {\doibase https://doi.org/10.1016/j.carbon.2021.01.154}
  {\bibfield  {journal} {\bibinfo  {journal} {Carbon}\ } (\bibinfo {year}
  {2021}),\ https://doi.org/10.1016/j.carbon.2021.01.154}\BibitemShut {NoStop}%
\bibitem [{\citenamefont {Agrait}\ \emph {et~al.}(1992)\citenamefont {Agrait},
  \citenamefont {Rodrigo},\ and\ \citenamefont {Vieira}}]{agr92}%
  \BibitemOpen
  \bibfield  {author} {\bibinfo {author} {\bibfnamefont {N.}~\bibnamefont
  {Agrait}}, \bibinfo {author} {\bibfnamefont {J.}~\bibnamefont {Rodrigo}}, \
  and\ \bibinfo {author} {\bibfnamefont {S.}~\bibnamefont {Vieira}},\
  }\href@noop {} {\bibfield  {journal} {\bibinfo  {journal} {Ultramicroscopy}\
  }\textbf {\bibinfo {volume} {42 -- 44, Part 1}},\ \bibinfo {pages} {177 }
  (\bibinfo {year} {1992})}\BibitemShut {NoStop}%
\bibitem [{\citenamefont {Schnedler}\ \emph
  {et~al.}(2015{\natexlab{a}})\citenamefont {Schnedler}, \citenamefont {Portz},
  \citenamefont {Weidlich}, \citenamefont {Dunin-Borkowski},\ and\
  \citenamefont {{Ph. Ebert}}}]{schnedler:2015b}%
  \BibitemOpen
  \bibfield  {author} {\bibinfo {author} {\bibfnamefont {M.}~\bibnamefont
  {Schnedler}}, \bibinfo {author} {\bibfnamefont {V.}~\bibnamefont {Portz}},
  \bibinfo {author} {\bibfnamefont {P.~H.}\ \bibnamefont {Weidlich}}, \bibinfo
  {author} {\bibfnamefont {R.~E.}\ \bibnamefont {Dunin-Borkowski}}, \ and\
  \bibinfo {author} {\bibnamefont {{Ph. Ebert}}},\ }\href {\doibase
  10.1103/PhysRevB.91.235305} {\bibfield  {journal} {\bibinfo  {journal} {Phys.
  Rev. B}\ }\textbf {\bibinfo {volume} {91}},\ \bibinfo {pages} {235305}
  (\bibinfo {year} {2015}{\natexlab{a}})}\BibitemShut {NoStop}%
\bibitem [{\citenamefont {Schnedler}\ \emph {et~al.}(2016)\citenamefont
  {Schnedler}, \citenamefont {Dunin-Borkowski},\ and\ \citenamefont {{Ph.
  Ebert}}}]{schnedler:2016a}%
  \BibitemOpen
  \bibfield  {author} {\bibinfo {author} {\bibfnamefont {M.}~\bibnamefont
  {Schnedler}}, \bibinfo {author} {\bibfnamefont {R.~E.}\ \bibnamefont
  {Dunin-Borkowski}}, \ and\ \bibinfo {author} {\bibnamefont {{Ph. Ebert}}},\
  }\href {\doibase 10.1103/PhysRevB.93.195444} {\bibfield  {journal} {\bibinfo
  {journal} {Phys. Rev. B}\ }\textbf {\bibinfo {volume} {93}},\ \bibinfo
  {pages} {195444} (\bibinfo {year} {2016})}\BibitemShut {NoStop}%
\bibitem [{\citenamefont {Rozplocha}\ \emph {et~al.}(2007)\citenamefont
  {Rozplocha}, \citenamefont {Patyk},\ and\ \citenamefont
  {Stankowski}}]{roz07}%
  \BibitemOpen
  \bibfield  {author} {\bibinfo {author} {\bibfnamefont {F.}~\bibnamefont
  {Rozplocha}}, \bibinfo {author} {\bibfnamefont {J.}~\bibnamefont {Patyk}}, \
  and\ \bibinfo {author} {\bibfnamefont {J.}~\bibnamefont {Stankowski}},\
  }\href@noop {} {\bibfield  {journal} {\bibinfo  {journal} {Acta Physica
  Polonica A}\ }\textbf {\bibinfo {volume} {112}} (\bibinfo {year} {2007})},\
  \Eprint
  {http://arxiv.org/abs/http://przyrbwn.icm.edu.pl/APP/PDF/112/a112z308.pdf}
  {http://przyrbwn.icm.edu.pl/APP/PDF/112/a112z308.pdf} \BibitemShut {NoStop}%
\bibitem [{\citenamefont {Dusari}\ \emph {et~al.}(2011)\citenamefont {Dusari},
  \citenamefont {Barzola-Quiquia}, \citenamefont {Esquinazi},\ and\
  \citenamefont {Garc\'ia}}]{dus11}%
  \BibitemOpen
  \bibfield  {author} {\bibinfo {author} {\bibfnamefont {S.}~\bibnamefont
  {Dusari}}, \bibinfo {author} {\bibfnamefont {J.}~\bibnamefont
  {Barzola-Quiquia}}, \bibinfo {author} {\bibfnamefont {P.}~\bibnamefont
  {Esquinazi}}, \ and\ \bibinfo {author} {\bibfnamefont {N.}~\bibnamefont
  {Garc\'ia}},\ }\href@noop {} {\bibfield  {journal} {\bibinfo  {journal}
  {Phys. Rev. B}\ }\textbf {\bibinfo {volume} {83}},\ \bibinfo {pages} {125402}
  (\bibinfo {year} {2011})}\BibitemShut {NoStop}%
\bibitem [{\citenamefont {Arndt}\ \emph {et~al.}(2009)\citenamefont {Arndt},
  \citenamefont {Spoddig}, \citenamefont {Esquinazi}, \citenamefont
  {Barzola-Quiquia}, \citenamefont {Dusari},\ and\ \citenamefont
  {Butz}}]{arn09}%
  \BibitemOpen
  \bibfield  {author} {\bibinfo {author} {\bibfnamefont {A.}~\bibnamefont
  {Arndt}}, \bibinfo {author} {\bibfnamefont {D.}~\bibnamefont {Spoddig}},
  \bibinfo {author} {\bibfnamefont {P.}~\bibnamefont {Esquinazi}}, \bibinfo
  {author} {\bibfnamefont {J.}~\bibnamefont {Barzola-Quiquia}}, \bibinfo
  {author} {\bibfnamefont {S.}~\bibnamefont {Dusari}}, \ and\ \bibinfo {author}
  {\bibfnamefont {T.}~\bibnamefont {Butz}},\ }\href@noop {} {\bibfield
  {journal} {\bibinfo  {journal} {Phys. Rev. B}\ }\textbf {\bibinfo {volume}
  {80}},\ \bibinfo {pages} {195402} (\bibinfo {year} {2009})}\BibitemShut
  {NoStop}%
\bibitem [{\citenamefont {Kaack}\ and\ \citenamefont {Fick}(1995)}]{kaa95}%
  \BibitemOpen
  \bibfield  {author} {\bibinfo {author} {\bibfnamefont {M.}~\bibnamefont
  {Kaack}}\ and\ \bibinfo {author} {\bibfnamefont {D.}~\bibnamefont {Fick}},\
  }\href {\doibase https://doi.org/10.1016/0039-6028(95)00758-X} {\bibfield
  {journal} {\bibinfo  {journal} {Surface Science}\ }\textbf {\bibinfo {volume}
  {342}},\ \bibinfo {pages} {111 } (\bibinfo {year} {1995})}\BibitemShut
  {NoStop}%
\bibitem [{\citenamefont {Krishnan}\ and\ \citenamefont
  {Jain}(1952)}]{krishnan:1952}%
  \BibitemOpen
  \bibfield  {author} {\bibinfo {author} {\bibfnamefont {K.}~\bibnamefont
  {Krishnan}}\ and\ \bibinfo {author} {\bibfnamefont {S.}~\bibnamefont
  {Jain}},\ }\href {\doibase 10.1038/169702c0} {\bibfield  {journal} {\bibinfo
  {journal} {Nature}\ }\textbf {\bibinfo {volume} {169}},\ \bibinfo {pages}
  {702} (\bibinfo {year} {1952})}\BibitemShut {NoStop}%
\bibitem [{\citenamefont {Rut'kov}\ \emph {et~al.}(2020)\citenamefont
  {Rut'kov}, \citenamefont {Afanas'eva},\ and\ \citenamefont
  {Gall}}]{rutkov:2020}%
  \BibitemOpen
  \bibfield  {author} {\bibinfo {author} {\bibfnamefont {E.}~\bibnamefont
  {Rut'kov}}, \bibinfo {author} {\bibfnamefont {E.}~\bibnamefont {Afanas'eva}},
  \ and\ \bibinfo {author} {\bibfnamefont {N.}~\bibnamefont {Gall}},\ }\href
  {\doibase 10.1016/j.diamond.2019.107576} {\bibfield  {journal} {\bibinfo
  {journal} {Diamond and Related Materials}\ }\textbf {\bibinfo {volume}
  {101}},\ \bibinfo {pages} {107576} (\bibinfo {year} {2020})}\BibitemShut
  {NoStop}%
\bibitem [{\citenamefont {Bono}\ and\ \citenamefont {Good}(1986)}]{bono:1986}%
  \BibitemOpen
  \bibfield  {author} {\bibinfo {author} {\bibfnamefont {J.}~\bibnamefont
  {Bono}}\ and\ \bibinfo {author} {\bibfnamefont {R.~H.}\ \bibnamefont
  {Good}},\ }\href {\doibase 10.1016/0039-6028(86)90243-8} {\bibfield
  {journal} {\bibinfo  {journal} {Surf. Sci.}\ }\textbf {\bibinfo {volume}
  {175}},\ \bibinfo {pages} {415} (\bibinfo {year} {1986})}\BibitemShut
  {NoStop}%
\bibitem [{\citenamefont {Feenstra}\ and\ \citenamefont
  {Stroscio}(1987)}]{feenstra:1987b}%
  \BibitemOpen
  \bibfield  {author} {\bibinfo {author} {\bibfnamefont {R.~M.}\ \bibnamefont
  {Feenstra}}\ and\ \bibinfo {author} {\bibfnamefont {J.~A.}\ \bibnamefont
  {Stroscio}},\ }\href@noop {} {\bibfield  {journal} {\bibinfo  {journal} {J.
  Vac. Sci. Technol. B}\ }\textbf {\bibinfo {volume} {5}},\ \bibinfo {pages}
  {923} (\bibinfo {year} {1987})}\BibitemShut {NoStop}%
\bibitem [{\citenamefont {Mu{\~n}oz}\ \emph {et~al.}(2013)\citenamefont
  {Mu{\~n}oz}, \citenamefont {Covaci},\ and\ \citenamefont {Peeters}}]{mun13}%
  \BibitemOpen
  \bibfield  {author} {\bibinfo {author} {\bibfnamefont {W.~A.}\ \bibnamefont
  {Mu{\~n}oz}}, \bibinfo {author} {\bibfnamefont {L.}~\bibnamefont {Covaci}}, \
  and\ \bibinfo {author} {\bibfnamefont {F.}~\bibnamefont {Peeters}},\ }\href
  {\doibase 10.1103/PhysRevB.87.134509} {\bibfield  {journal} {\bibinfo
  {journal} {Phys. Rev. B}\ }\textbf {\bibinfo {volume} {87}},\ \bibinfo
  {pages} {134509} (\bibinfo {year} {2013})}\BibitemShut {NoStop}%
\bibitem [{\citenamefont {Xu}\ \emph {et~al.}(2015)\citenamefont {Xu},
  \citenamefont {Yin}, \citenamefont {Qiao}, \citenamefont {Bai}, \citenamefont
  {Nie},\ and\ \citenamefont {He}}]{xu15}%
  \BibitemOpen
  \bibfield  {author} {\bibinfo {author} {\bibfnamefont {R.}~\bibnamefont
  {Xu}}, \bibinfo {author} {\bibfnamefont {L.-J.}\ \bibnamefont {Yin}},
  \bibinfo {author} {\bibfnamefont {J.-B.}\ \bibnamefont {Qiao}}, \bibinfo
  {author} {\bibfnamefont {K.-K.}\ \bibnamefont {Bai}}, \bibinfo {author}
  {\bibfnamefont {J.-C.}\ \bibnamefont {Nie}}, \ and\ \bibinfo {author}
  {\bibfnamefont {L.}~\bibnamefont {He}},\ }\href {\doibase
  10.1103/PhysRevB.91.035410} {\bibfield  {journal} {\bibinfo  {journal} {Phys.
  Rev. B}\ }\textbf {\bibinfo {volume} {91}},\ \bibinfo {pages} {035410}
  (\bibinfo {year} {2015})}\BibitemShut {NoStop}%
\bibitem [{\citenamefont {Wang}\ \emph {et~al.}(2018)\citenamefont {Wang},
  \citenamefont {Shi}, \citenamefont {Zakharov}, \citenamefont {Syv\"aj\"arvi},
  \citenamefont {Yakimova}, \citenamefont {Uhrberg},\ and\ \citenamefont
  {Sun}}]{wan18}%
  \BibitemOpen
  \bibfield  {author} {\bibinfo {author} {\bibfnamefont {W.}~\bibnamefont
  {Wang}}, \bibinfo {author} {\bibfnamefont {Y.}~\bibnamefont {Shi}}, \bibinfo
  {author} {\bibfnamefont {A.~A.}\ \bibnamefont {Zakharov}}, \bibinfo {author}
  {\bibfnamefont {M.}~\bibnamefont {Syv\"aj\"arvi}}, \bibinfo {author}
  {\bibfnamefont {R.}~\bibnamefont {Yakimova}}, \bibinfo {author}
  {\bibfnamefont {R.~I.~G.}\ \bibnamefont {Uhrberg}}, \ and\ \bibinfo {author}
  {\bibfnamefont {J.}~\bibnamefont {Sun}},\ }\href {\doibase
  10.1021/acs.nanolett.8b02530} {\bibfield  {journal} {\bibinfo  {journal}
  {Nano Letters}\ }\textbf {\bibinfo {volume} {18}},\ \bibinfo {pages} {5862}
  (\bibinfo {year} {2018})},\ \Eprint
  {http://arxiv.org/abs/https://doi.org/10.1021/acs.nanolett.8b02530}
  {https://doi.org/10.1021/acs.nanolett.8b02530} \BibitemShut {NoStop}%
\bibitem [{\citenamefont {Heikkil\"a}\ and\ \citenamefont
  {Volovik}(2016)}]{hei16}%
  \BibitemOpen
  \bibfield  {author} {\bibinfo {author} {\bibfnamefont {T.}~\bibnamefont
  {Heikkil\"a}}\ and\ \bibinfo {author} {\bibfnamefont {G.~E.}\ \bibnamefont
  {Volovik}},\ }\enquote {\bibinfo {title} {Flat bands as a route to
  high-temperature superconductivity in graphite},}\ \ (\bibinfo  {publisher}
  {P. Esquinazi (ed.), Springer International Publishing AG Switzerland},\
  \bibinfo {year} {2016})\ pp.\ \bibinfo {pages} {123--144}\BibitemShut
  {NoStop}%
\bibitem [{\citenamefont {Pamuk}\ \emph {et~al.}(2017)\citenamefont {Pamuk},
  \citenamefont {Baima}, \citenamefont {Mauri},\ and\ \citenamefont
  {Calandra}}]{pam17}%
  \BibitemOpen
  \bibfield  {author} {\bibinfo {author} {\bibfnamefont {B.}~\bibnamefont
  {Pamuk}}, \bibinfo {author} {\bibfnamefont {J.}~\bibnamefont {Baima}},
  \bibinfo {author} {\bibfnamefont {F.}~\bibnamefont {Mauri}}, \ and\ \bibinfo
  {author} {\bibfnamefont {M.}~\bibnamefont {Calandra}},\ }\href {\doibase
  10.1103/PhysRevB.95.075422} {\bibfield  {journal} {\bibinfo  {journal} {Phys.
  Rev. B}\ }\textbf {\bibinfo {volume} {95}},\ \bibinfo {pages} {075422}
  (\bibinfo {year} {2017})}\BibitemShut {NoStop}%
\bibitem [{\citenamefont {Ojaj\"arvi}\ \emph {et~al.}(2018)\citenamefont
  {Ojaj\"arvi}, \citenamefont {Hyart}, \citenamefont {Silaev},\ and\
  \citenamefont {Heikkil\"a}}]{oja18}%
  \BibitemOpen
  \bibfield  {author} {\bibinfo {author} {\bibfnamefont {R.}~\bibnamefont
  {Ojaj\"arvi}}, \bibinfo {author} {\bibfnamefont {T.}~\bibnamefont {Hyart}},
  \bibinfo {author} {\bibfnamefont {M.~A.}\ \bibnamefont {Silaev}}, \ and\
  \bibinfo {author} {\bibfnamefont {T.~T.}\ \bibnamefont {Heikkil\"a}},\ }\href
  {\doibase 10.1103/PhysRevB.98.054515} {\bibfield  {journal} {\bibinfo
  {journal} {Phys. Rev. B}\ }\textbf {\bibinfo {volume} {98}},\ \bibinfo
  {pages} {054515} (\bibinfo {year} {2018})}\BibitemShut {NoStop}%
\bibitem [{\citenamefont {Hyart}\ \emph {et~al.}(2018)\citenamefont {Hyart},
  \citenamefont {Ojaj\"arvi},\ and\ \citenamefont {Heikkil\"a}}]{hya18}%
  \BibitemOpen
  \bibfield  {author} {\bibinfo {author} {\bibfnamefont {T.}~\bibnamefont
  {Hyart}}, \bibinfo {author} {\bibfnamefont {R.}~\bibnamefont {Ojaj\"arvi}}, \
  and\ \bibinfo {author} {\bibfnamefont {T.}~\bibnamefont {Heikkil\"a}},\
  }\href {https://doi.org/10.1007/s10909-017-1846-3} {\bibfield  {journal}
  {\bibinfo  {journal} {J. Low Temp Phys}\ }\textbf {\bibinfo {volume} {191}},\
  \bibinfo {pages} {35} (\bibinfo {year} {2018})},\ \Eprint
  {http://arxiv.org/abs/https://doi.org/10.1007/s10909-017-1846-3}
  {https://doi.org/10.1007/s10909-017-1846-3} \BibitemShut {NoStop}%
\bibitem [{\citenamefont {Schnedler}\ \emph
  {et~al.}(2015{\natexlab{b}})\citenamefont {Schnedler}, \citenamefont {Portz},
  \citenamefont {Eisele}, \citenamefont {Dunin-Borkowski},\ and\ \citenamefont
  {{Ph. Ebert}}}]{schnedler:2015a}%
  \BibitemOpen
  \bibfield  {author} {\bibinfo {author} {\bibfnamefont {M.}~\bibnamefont
  {Schnedler}}, \bibinfo {author} {\bibfnamefont {V.}~\bibnamefont {Portz}},
  \bibinfo {author} {\bibfnamefont {H.}~\bibnamefont {Eisele}}, \bibinfo
  {author} {\bibfnamefont {R.~E.}\ \bibnamefont {Dunin-Borkowski}}, \ and\
  \bibinfo {author} {\bibnamefont {{Ph. Ebert}}},\ }\href {\doibase
  10.1103/PhysRevB.91.205309} {\bibfield  {journal} {\bibinfo  {journal} {Phys.
  Rev. B}\ }\textbf {\bibinfo {volume} {91}},\ \bibinfo {pages} {205309}
  (\bibinfo {year} {2015}{\natexlab{b}})}\BibitemShut {NoStop}%
\bibitem [{\citenamefont {Portz}\ \emph {et~al.}(2017)\citenamefont {Portz},
  \citenamefont {Schnedler}, \citenamefont {Lymperakis}, \citenamefont
  {Neugebauer}, \citenamefont {Eisele}, \citenamefont {Carlin}, \citenamefont
  {Butt{\'e}}, \citenamefont {Grandjean}, \citenamefont {Dunin-Borkowski},\
  and\ \citenamefont {Ebert}}]{portz:2017}%
  \BibitemOpen
  \bibfield  {author} {\bibinfo {author} {\bibfnamefont {V.}~\bibnamefont
  {Portz}}, \bibinfo {author} {\bibfnamefont {M.}~\bibnamefont {Schnedler}},
  \bibinfo {author} {\bibfnamefont {L.}~\bibnamefont {Lymperakis}}, \bibinfo
  {author} {\bibfnamefont {J.}~\bibnamefont {Neugebauer}}, \bibinfo {author}
  {\bibfnamefont {H.}~\bibnamefont {Eisele}}, \bibinfo {author} {\bibfnamefont
  {J.-F.}\ \bibnamefont {Carlin}}, \bibinfo {author} {\bibfnamefont
  {R.}~\bibnamefont {Butt{\'e}}}, \bibinfo {author} {\bibfnamefont
  {N.}~\bibnamefont {Grandjean}}, \bibinfo {author} {\bibfnamefont {R.~E.}\
  \bibnamefont {Dunin-Borkowski}}, \ and\ \bibinfo {author} {\bibfnamefont
  {P.}~\bibnamefont {Ebert}},\ }\href {\doibase 10.1063/1.4973765} {\bibfield
  {journal} {\bibinfo  {journal} {Appl. Phys. Lett.}\ }\textbf {\bibinfo
  {volume} {110}},\ \bibinfo {pages} {022104} (\bibinfo {year}
  {2017})}\BibitemShut {NoStop}%
\bibitem [{\citenamefont {Spemann}\ and\ \citenamefont
  {Esquinazi}(2016)}]{chap3}%
  \BibitemOpen
  \bibfield  {author} {\bibinfo {author} {\bibfnamefont {D.}~\bibnamefont
  {Spemann}}\ and\ \bibinfo {author} {\bibfnamefont {P.}~\bibnamefont
  {Esquinazi}},\ }\enquote {\bibinfo {title} {Evidence for magnetic order in
  graphite from magnetization and transport measurements},}\ \ (\bibinfo
  {publisher} {P. Esquinazi (ed.), Springer International Publishing AG
  Switzerland},\ \bibinfo {year} {2016})\ pp.\ \bibinfo {pages}
  {45--76}\BibitemShut {NoStop}%
\end{thebibliography}
%

\end{document}